\newcommand*{\ATLASLATEXPATH}{}
\documentclass[PAPER, UKenglish, texlive=2016]{\ATLASLATEXPATH atlasdoc}
\pdfoutput=1
\usepackage{\ATLASLATEXPATH atlaspackage}
\usepackage{\ATLASLATEXPATH atlasbiblatex}

\usepackage{\ATLASLATEXPATH atlasphysics}
\graphicspath{{logos/}{figures/}}

\usepackage{authblk}
\author[1]{D.~Calvet}
\author[1]{S.~Calvet}
\author[1]{R.~Chadelas}
\author[1,3]{D.~Cinca}
\author[1,4]{Ph.~Grenier}
\author[1]{Ph.~Gris}
\author[1]{P.~Lafarguette}
\author[1]{D.~Lambert}
\author[1]{M.~Marjanovi\'c}
\author[2]{L.F.~Oleiro~Seabra}
\author[2]{F.M.~Pedro~Martins}
\author[2,5]{J.B.~Pena~Madeira~Gouveia~De~Campos}
\author[1]{S.M.~Romano~Saez}
\author[1]{Ph.~Rosnet}
\author[1]{C.~Santoni}
\author[1,6]{L.~Val\'ery}
\author[1]{F.~Vazeille}
\affil[1]{Universit\'e Clermont Auvergne, CNRS/IN2P3, LPC, Clermont-Ferrand}
\affil[2]{Laborat\'orio de Instrumenta\c{c}\~ao e F\'\i sica Experimental de Part\'\i culas - LIP, Lisboa}
\affil[3]{now at Technische Universitaet Dortmund}
\affil[4]{now at SLAC National Accelerator Laboratory, Stanford CA}
\affil[5]{now at LARSyS, Instituto Superior T\'ecnico, univ. Lisboa}
\affil[6]{now at DESY, Hamburg}

\AtlasAbstract{%

This article documents the characteristics of the high voltage (HV) system of the hadronic calorimeter TileCal of the ATLAS experiment. Such a system is suitable to supply reliable power distribution into particles physics detectors using a large number of PhotoMultiplier Tubes (PMTs). Measurements performed during the 2015 and 2016 data taking periods of the ATLAS detector show that its performance, in terms of stability and noise, fits the specifications. In particular, almost all the PMTs show a voltage instability smaller than 0.5 V corresponding to a gain stability better than 0.5\%. A small amount of channels was found not working correctly. To diagnose the origin of such defects, the results of the HV measurements were compared to those obtained using a Laser system. The analysis shows that less than 0.2\% of the about 10 thousand HV channels were malfunctioning.

}

\hypersetup{pdftitle={ATLAS document},pdfauthor={The ATLAS Collaboration}}

\newcommand{\HVin}{\ensuremath{\mathrm{HV}_{\mathrm{in}}}}

\newcommand{\HVout}{\ensuremath{\mathrm{HV}_{\mathrm{out}}^i}}
\newcommand{\HVnom}{\ensuremath{\mathrm{HV}_{\mathrm{nom}}^i}}
\newcommand{\HVmeas}{\ensuremath{\mathrm{HV}_{\mathrm{meas}}^i}}
\newcommand{\HVleft}{\ensuremath{\mathrm{HV}_{\mathrm{in}}^{\mathrm{left}}}}
\newcommand{\HVright}{\ensuremath{\mathrm{HV}_{\mathrm{in}}^{\mathrm{right}}}}
\newcommand{\HVleftright}{\ensuremath{\mathrm{HV}_{\mathrm{in}}^{\mathrm{left/right}}}}
\newcommand{\HVmax}{\ensuremath{\mathrm{HV}_{\mathrm{max}}^i}}
\newcommand{\DAC}{\ensuremath{\mathrm{DAC}^i}}
\newcommand{\ADC}{\ensuremath{\mathrm{ADC}^i}}
\newcommand{\DeltaHV}{\ensuremath{\Delta\mathrm{HV}^i}}
\newcommand{\AvDeltaHV}{\ensuremath{\overline{\Delta\mathrm{HV}}^i}}
\newcommand{\mudhvi}{\ensuremath{\mu_{\Delta\mathrm{HV}}^i}}
\newcommand{\sigdhvi}{\ensuremath{\sigma_{\Delta\mathrm{HV}}^i}}
\newcommand{\mudhvp}{\ensuremath{\mu_{\Delta\mathrm{HV}}^{part}}}
\newcommand{\sigdhvp}{\ensuremath{\sigma_{\Delta\mathrm{HV}}^{part}}}

\newcommand{\Gi}{\ensuremath{G^i}}
\newcommand{\tref}{\ensuremath{t_{\mathrm{ref}}}}
\newcommand{\GHV}{\ensuremath{G^i_{\mathrm{HV}}}}
\newcommand{\GLaser}{\ensuremath{G^i_{\mathrm{Laser}}}}

\begin{document}
%\linenumbers
%
\title{The High Voltage distribution system of the ATLAS Tile Calorimeter and 
its performance during data taking}

\maketitle

\tableofcontents

% List of contributors - print here or after the Bibliography.
%\PrintAtlasContribute{0.30}
%\clearpage

%-------------------------------------------------------------------------------
%\section*{Introduction}
\section{Introduction}
\label{sec:intro}
%-------------------------------------------------------------------------------
The High Voltage (HV) distribution system described in this article has been developed to supply each of the about 10000 PhotoMultiplier Tubes (PMTs) of the ATLAS Tile Calorimeter (TileCal)~\cite{tdr_tile,paper_tile}, with a voltage ranging from  approximately 500 to 900~V. TileCal is the central hadronic calorimeter of the ATLAS~\cite{PERF-2007-01} experiment at the Large Hadron Collider (LHC)~\cite{Evans:2008zzb} of CERN. It is a sampling calorimeter whose operation is based on the detection of scintillation light using PMTs. It plays an important role for the precise reconstruction of the kinematics of the physics events, because about  30~\% of the total energy of jets produced in the proton collisions is deposited in TileCal. In order to keep stable the energy  measurements, the PMTs must be supplied with a very stable HV. As discussed in Section~\ref{sec:HVstability}, the relation between the gain $G$ of a PMT and the applied high voltage HV is  given by $G\propto\mathrm{HV}^\beta$,   
%\begin{equation}
%G= cst \times ( \mathrm{HV} )^{\beta} 
%\label{eq:Gbeta}                                       
%\end{equation}
where $\beta$ is specific to each PMT. Any variation of the high voltage would induce a gain variation deteriorating the measurement of the energy deposited by the particles in the calorimeter cells.

%The high voltage regulation system described here has been realized to supply the photo-multipliers of the tile hadron calorimeter TileCal~\cite{tdr_tile},~\cite{PERF-2007-01} of the ATLAS detector~\cite{PERF-2007-01}. 
A short description of the calorimeter is reported in Section~\ref{sec:Tile}. The hardware and the online monitoring software of the HV system are described in Section~\ref{sec:HV}. The tolerance to radiation of the system appears in Section~\ref{sec:tolrad} and the characteristics and performance of the system in Section~\ref{sec:characteristic}. The results of an analysis of faulty boards are discussed in Section~\ref{sec:faulty}. The measurements performed during data taking concerning the most important feature, the stability, are presented in Section~\ref{sec:HVstability}. Finally, conclusions are drawn in Section~\ref{conclusions}.

%-------------------------------------------------------------------------------
\section{The Tile Calorimeter}
\label{sec:Tile}
%-------------------------------------------------------------------------------
The TileCal calorimeter is described extensively in several publications~\cite{tdr_tile,paper_tile,tilerun1},
only the information relevant to the HV distribution system is given here.

\begin{figure}[t]
% Use the relevant command for your figure-insertion program
% to insert the figure file.
\centering
  \setlength{\unitlength}{0.1\textwidth}
  \begin{picture}(10,6.5)
    \put(0,0){\includegraphics[width=0.6\textwidth]{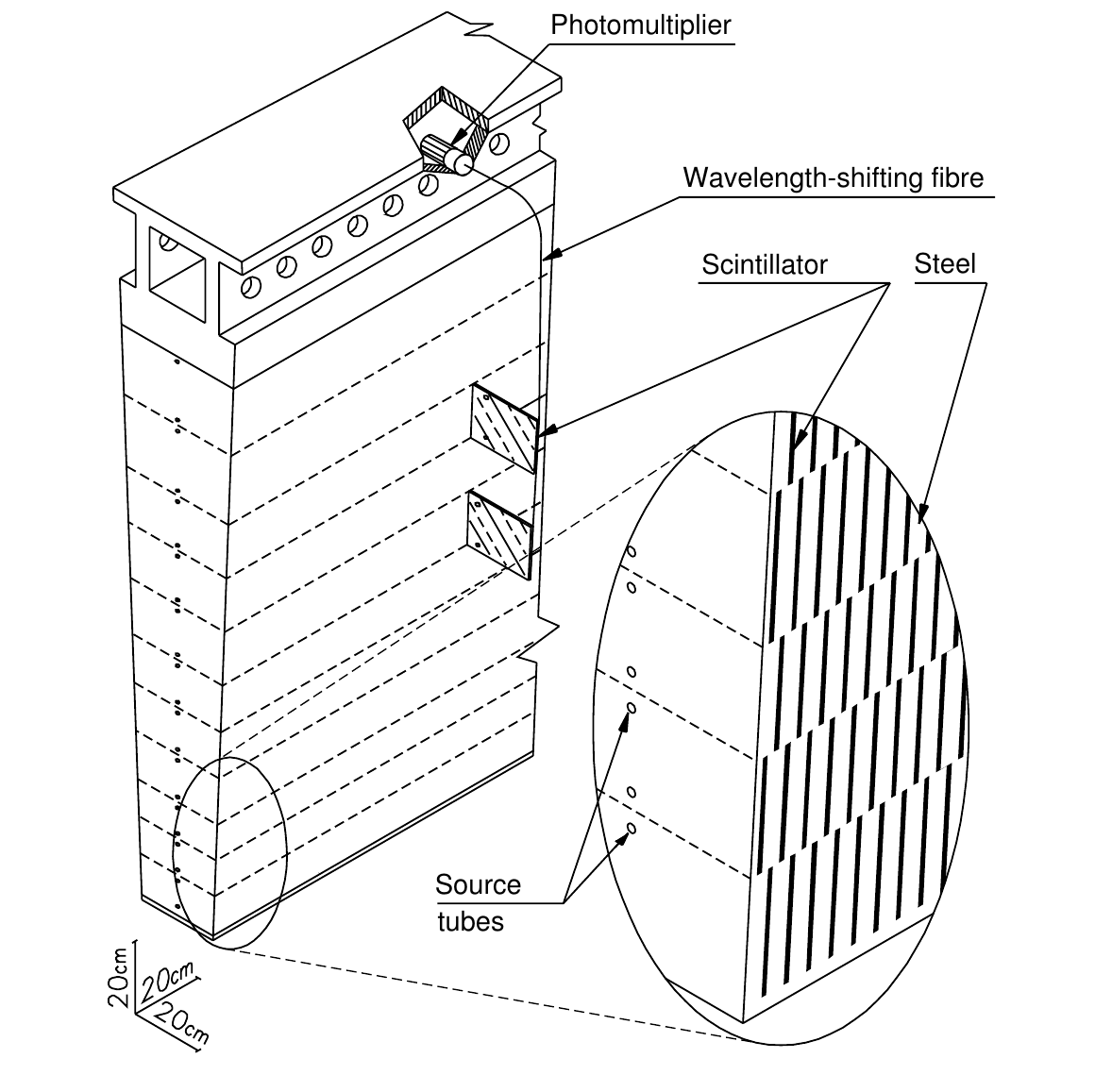}}
    \put(1,4.8){\scriptsize Girder}
  \end{picture}  
\caption{Drawing of a TileCal module showing the structure of the iron and scintillator tiles and the method of light collection by wavelength shifting fibres to the photomultipliers. The holes for the radioactive source tubes that traverse all the eleven radial tile-rows parallel to the colliding beams are also shown~\cite{tdr_tile,PERF-2007-01}.}
\label{fig:module}       % Give a unique label
\end{figure}

The calorimeter is a large cylinder with the LHC beam as axis, mechanically divided into three parts: the central part, called Long Barrel (LB) and two end-caps, the Extended Barrels (EBA and EBC). Each of these three barrels is assembled from 64 wedge-shaped modules (see Fig.~\ref{fig:module}), labeled from 01 to 64. The modules have a periodical structure of iron plates and scintillating tiles, the tiles being perpendicular to the beam axis. Wavelength shifting fibres transmit light produced in the tiles to 9852 photomultipliers~\cite{PMT}: except in some exceptionnal cases, each tile is read out by two fibres, from both sides (left and right). In each module a three-dimensional cell structure is defined by grouping several optical fibres connected to the same PMT~\cite{Abdallah:2009zza}: two photomultipliers (for left and right fibres) read-out each cell and the signals of the two PMTs are summed up to provide the cell response (except for a few special cases, when a single PMT is connected to the cell).

The photomultipliers and the associated front-end electronics are located in a super-drawer 2.60~m long, composed of an external and an internal drawer, inserted in the girder positioned on the external part of the calorimeter modules (see Fig.~\ref{fig:module}). Each EB module is read-out by one super-drawer, while two super-drawers are needed for each LB module. Therefore, in total, 256 super-drawers are inserted in the 192 modules.  In each super-drawer can be installed a maximum of 48 PMTs (24 in each of the two drawers that constitute a super-drawer). In EB modules, most of the super-drawers contain only 32 PMTs while 45 PMTs are present in LB super-drawers. Attached to each PMT, a voltage divider~\cite{tdr_tile,PERF-2007-01} provides electric connections for the supplied high voltage and signal output. 

From the electronics point of view, the super-drawers are grouped in four partitions: the EBA and EBC partitions, corresponding to the EBA and EBC mechanical barrels, and the LBA and LBC partitions, corresponding each to half of the LB central barrel. LBA (LBC) super-drawers are close to the EBA (EBC) barrel.

%-------------------------------------------------------------------------------
\section{The High Voltage Distribution System}
\label{sec:HV}
%-------------------------------------------------------------------------------
This section contains a detailed description of the hardware of the HV system, as well as its associated online monitoring infrastructure.

The main specifications of the HV distribution system are:
\begin{itemize}
\item To be able to set a different high voltage value to each of the 9852 TileCal PMTs with a limited number of HV power supplies (see Section~\ref{subsec:HVsource});
\item the applied high voltage must be stable, with a spread smaller than 0.5~V, thus ensuring a stability of the PMT gain better than 0.5\%, and a ripple smaller than 20 mV peak to peak;
\item the sensitivity to temperature variations must be smaller than 0.2~V/$^\circ$C and it must be insensitive to humidity values up to 60\% (see Section~\ref{subsubsec:humidity});
\item it must be insensitive to magnetic fields up to 0.1~T;
\item it must be able to deliver a current up to 350~$\mu$A/PMT;
\item when absolutely necessary to switch off some PMTs, the impact on the cell energy measurement must be limited, for example by keeping one PMT out of the two that are reading-out the cell.
\end{itemize}
Taking into account these specifications, the system has been designed with a single HV
power supply per super-drawer powering up to 48 PMTs. Each single HV power supply is then able to deliver up to 15.8~mA. The voltage adjustment for each PMT is performed
by an individual regulation loop described in Section~\ref{subsec:HVopto}. The emergency switches
control the 12~left or the 12~right PMTs of a super-drawer (see Section~\ref{sec:Tile}) in a way that, in case of problems, it is possible to obtain a measurement, although less precise,  of the energy deposited in the affected cells using the signal measured in the operational PMTs. In addition, the HV distribution system includes several temperature probes to monitor the super-drawer temperature (see Section~\ref{subsec:temp-mon}).

The channels are monitored and adjusted individually using the Detector Control System (DCS)~\cite{Jenni:616089}, discussed in Section~\ref{subsec:DCS}. 

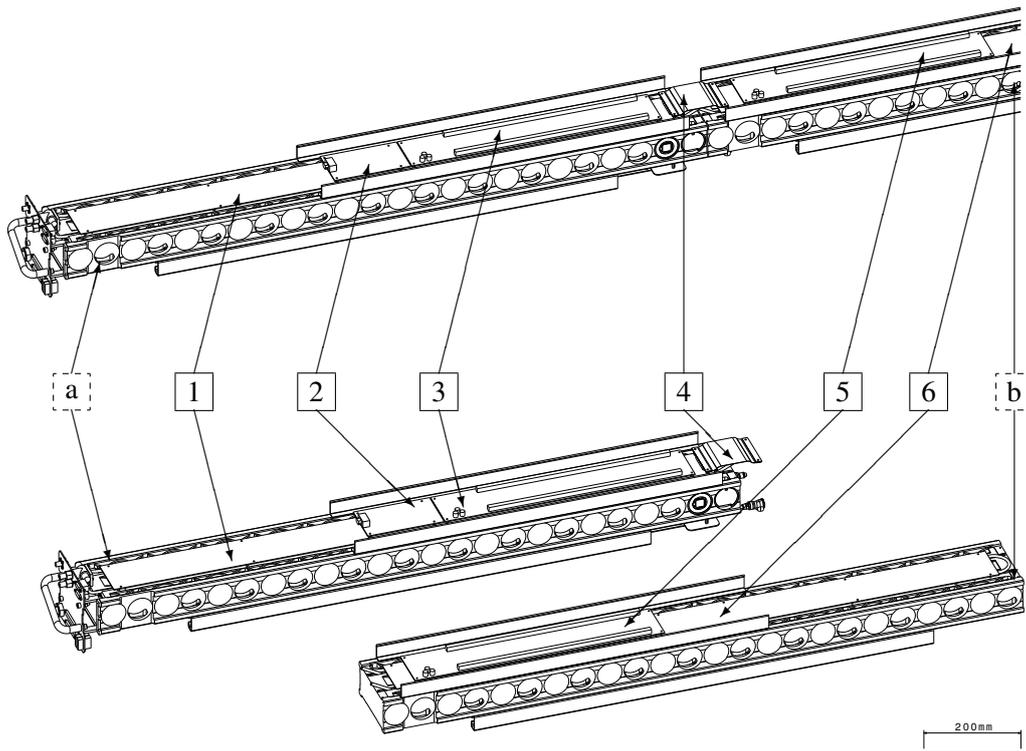
\begin{figure}[t]
  % Use the relevant command for your figure-insertion program
  % to insert the figure file.
  \centering
  \setlength{\unitlength}{0.1\textwidth}
  \begin{picture}(10,6.5)
    \put(0,0){\includegraphics[width=\textwidth]{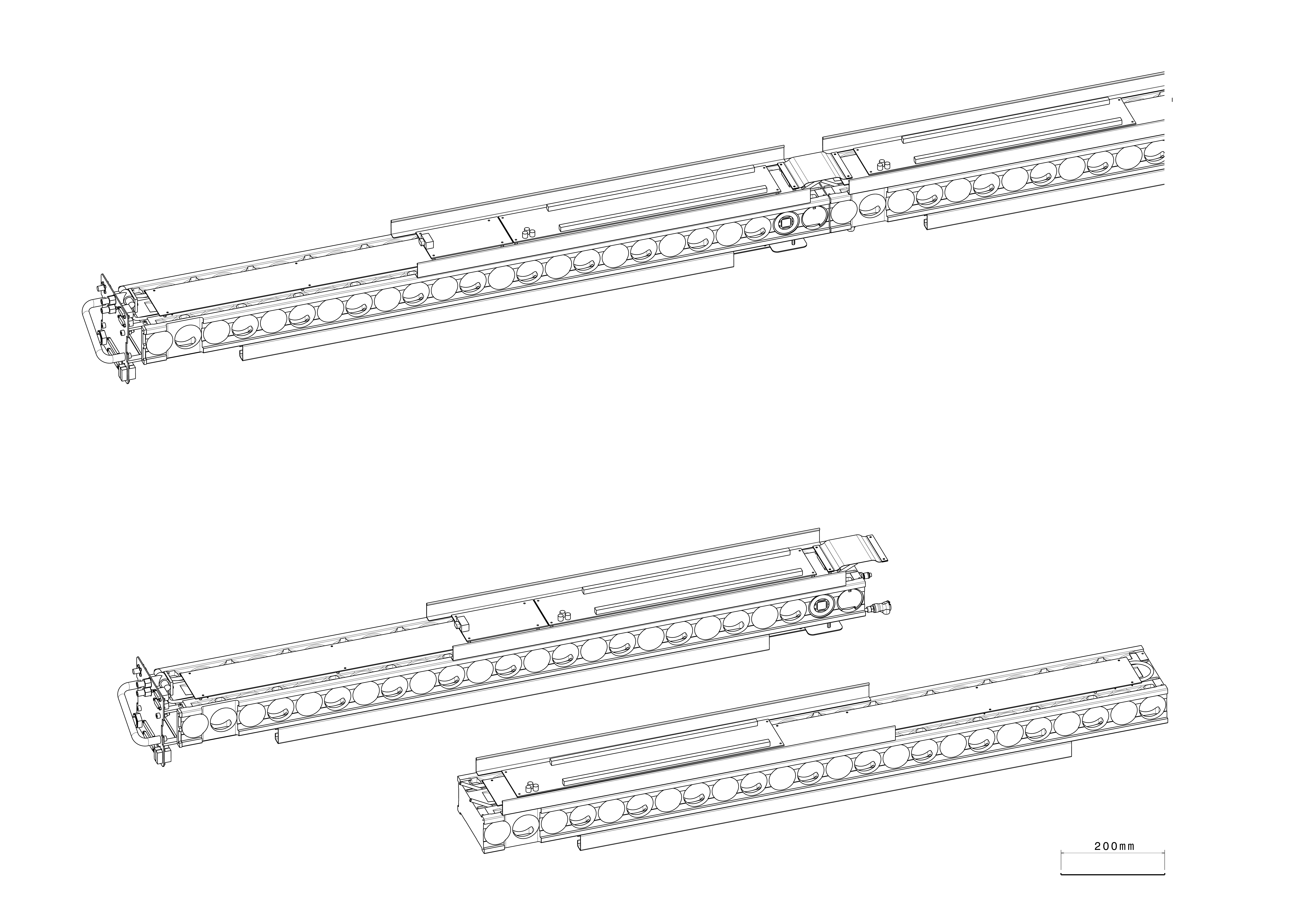}}
%    \multiput(0,0)(0,1){10}{\line(1,0){10}}
%    \multiput(0,0)(1,0){10}{\line(0,1){10}}
    \put(1,3.2){\dashbox{0.05}(0.3,0.3){a}}
    \put(1.15,3.2){\vector(1,-4){0.31}}
    \put(1.15,3.5){\vector(1,4){0.23}}
    \put(2,3.2){\framebox(0.3,0.3){1}}
    \put(2.15,3.2){\vector(1,-4){0.31}}
    \put(2.15,3.5){\vector(1,4){0.37}}
    \put(3,3.2){\framebox(0.3,0.3){2}}
    \put(3.15,3.2){\vector(1,-1){0.8}}
    \put(3.15,3.5){\vector(1,4){0.43}}
    \put(4,3.2){\framebox(0.3,0.3){3}}
    \put(4.15,3.2){\vector(1,-4){0.2}}
    \put(4.15,3.5){\vector(1,4){0.5}}
    \put(6,3.2){\framebox(0.3,0.3){4}}
    \put(6.15,3.2){\vector(1,-1){0.4}}
    \put(6.15,3.5){\vector(0,1){2.3}}
    \put(7.3,3.2){\framebox(0.3,0.3){5}}
    \put(7.45,3.2){\vector(-1,-1){1.78}}
    \put(7.45,3.5){\vector(1,4){0.66}}
    \put(8,3.2){\framebox(0.3,0.3){6}}
    \put(8.15,3.2){\vector(-1,-1){1.7}}
    \put(8.15,3.5){\vector(1,4){0.68}}
    \put(8.7,3.2){\dashbox{0.05}(0.3,0.3){b}}
    \put(8.85,3.2){\vector(0,-1){1.4}}
    \put(8.85,3.5){\vector(0,1){2.4}}
  \end{picture}  
  \caption{Sketch of a partial super-drawer showing the external drawer (a) and the internal drawer (b). The two unconnected drawers are shown on the bottom. The HV components are indicated: the  external HVbus board (1), the HVmicro board (2), the external HVopto board (3), the HVflex (4), the internal HVopto board (5) and the  internal HVbus board (6).}
  \label{fig:diagram}       % Give a unique label
\end{figure}
\subsection{General description of the HV system}
\label{subsec:general}
\begin{figure}[p]
% Use the relevant command for your figure-insertion program
% to insert the figure file.
\centering
\includegraphics[width=\textwidth,clip]{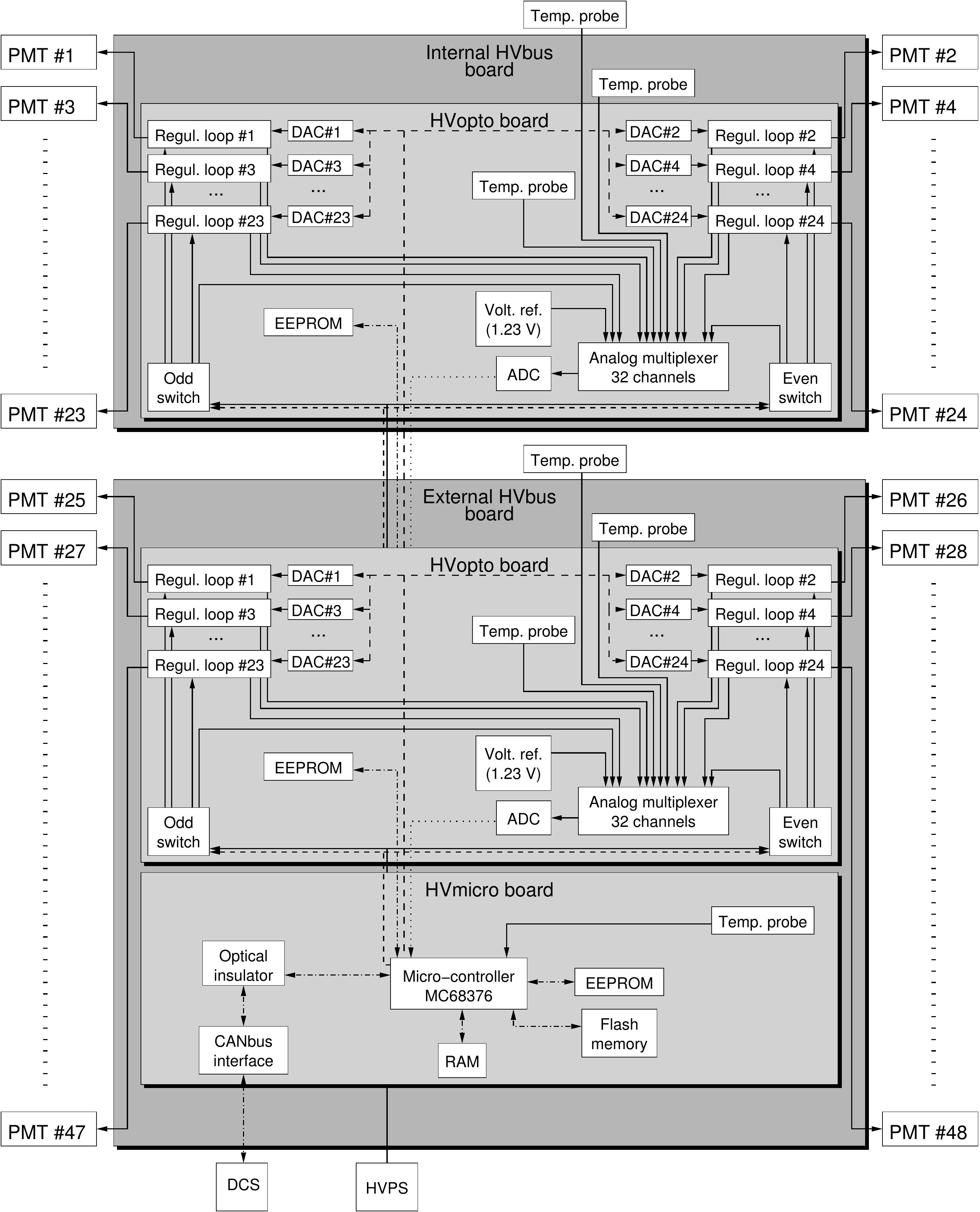}
\caption{Scheme of the HV distribution system. In the HVopto board the odd  and even components correspond to the left and right PMTs respectively discussed in the text.}
\label{fig:scheme}       % Give a unique label
\end{figure}

The HV distribution system is made of two parts: a front-end system for the individual regulation and distribution of the voltages and an external HV power supply~\cite{HV_power} localized in the ATLAS technical cavern USA15~\cite{PERF-2007-01}. As shown in  Fig.~\ref{fig:diagram} the boards are placed at one surface of the super-drawer.  The front-end data readout electronics, not visible in the figure, is hosted in the other side. The HV electronics for one super-drawer consists of:
\begin{itemize} 
\item One HVmicro board (166 $\times$ 95 mm$^2$) that hosts a micro-controller (Motorola MC68376) allowing driving the HVopto boards.
\item Two HVopto boards (544 $\times$ 95 mm$^2$), one internal and one external. Each of them provides the individual regulation of the 24 PMTs of a drawer.
\item Two long HVbus boards (1312 $\times$ 92 mm$^2$), one internal and one external, distribute the adjusted voltages to the PMTs. The HVmicro board and one of the two HVopto boards are installed on the HVbus board localized in the external drawer. The other HVopto board is placed on the HVbus board of the internal drawer. 
\item A short flexible bus (HVflex) links the external and the internal HVbus boards.
\end{itemize}
The scheme of the HV distribution system is shown in Fig.~\ref{fig:scheme}. The boards are supplied with $+5$~V, $+15$~V and $-15$~V by an external power supply~\cite{LV_power} located in the girder, close to the external drawer. A voltage of $-5$~V is locally generated in the HVopto boards using regulators.

\subsection{Temperature monitoring}
\label{subsec:temp-mon}
The TileCal super-drawers are water-cooled using a circuit below the atmospheric pressure to avoid leakages. As already mentioned the HV distribution system allows also monitoring of the temperature inside each super-drawer. Seven temperature probes are  monitored by the HVmicro board. These probes are temperature transducers (AD590) that produce an output current proportional to temperature. The probes are integrated on the HVmicro board, the HVopto boards and the HVbus boards (see Fig.~\ref{fig:scheme}). Two external probes are connected to the HVopto boards, one is located close to a PMT and the other is close to the interface board of the readout system placed on the other side of the super-drawer. The temperature monitoring accuracy is 0.1$^\circ$C.

\subsection{HV power supply}
\label{subsec:HVsource}
The high voltage power supply consists of 16 custom-made High Voltage Power Supplies (HVPS)~\cite{HV_power} independent units. Each unit has 16 output channels and each channel provides the high voltage, {\HVin}, to a super-drawer. Two {\HVin} values equal to $-830$ V and $-950$ V with a nominal consumption of 11~mA (maximal output current is 15.8~mA for LB's and 11.2~mA for EB's) can be provided. The choice of the two possible HV values for a super-drawer is performed on the base of the scintillating tile quality (two different manufacturers produced the scintillating polystyrene material~\cite{Abdallah:2009zza}), the radiation aging of the tiles in the module, the quantum efficiency of the connected PMTs~\cite{tdr_tile} and their nominal high voltages~\cite{PMT}. The {\HVin} values can be set and monitored remotely using the DCS system discussed in Section~\ref{subsec:DCS}.

\subsection{HVmicro board}
\label{subsec:HVmicro}
The HVmicro board of each super-drawer allows to set the appropriate high voltage to each of the 48 channels (one value per PMT), and also to monitor the applied values for all the PMTs with a frequency of approximately 2~Hz. The micro-controller works with a 20~MHz clock. Its program, written in C, performs the following tasks:
\begin{itemize}
\item setting of the correct HV value for each channel, using a Digital to Analog Converter (DAC) located on the corresponding HVopto board;
\item monitoring of the 48 applied HV values, using one Analog to Digital Converter (ADC) located on each HVopto board;
\item monitoring of the input high voltage (\HVin) for each group of 12 PMTs, the seven input low voltages ($+5$~V, $-15$~V, $+15$~V, four $-5$~V), the seven temperature probe values and two reference voltages (see Section~\ref{subsec:HVopto} );
\item conversion of the DAC and ADC values to and from physical values (voltages, temperatures) (see Section~\ref{subsec:conversion});
\item control of the four emergency switches, located on the HVopto boards;
\item continuous correction against Single Event Effect (SEE) discussed in Section~\ref{sec:tolrad};
\item communication with the ATLAS DCS using the CANbus network~\cite{Jenni:616089}.
\end{itemize}

\begin{figure}[t]
% Use the relevant command for your figure-insertion program
% to insert the figure file.
\centering
\includegraphics[width=\textwidth,clip]{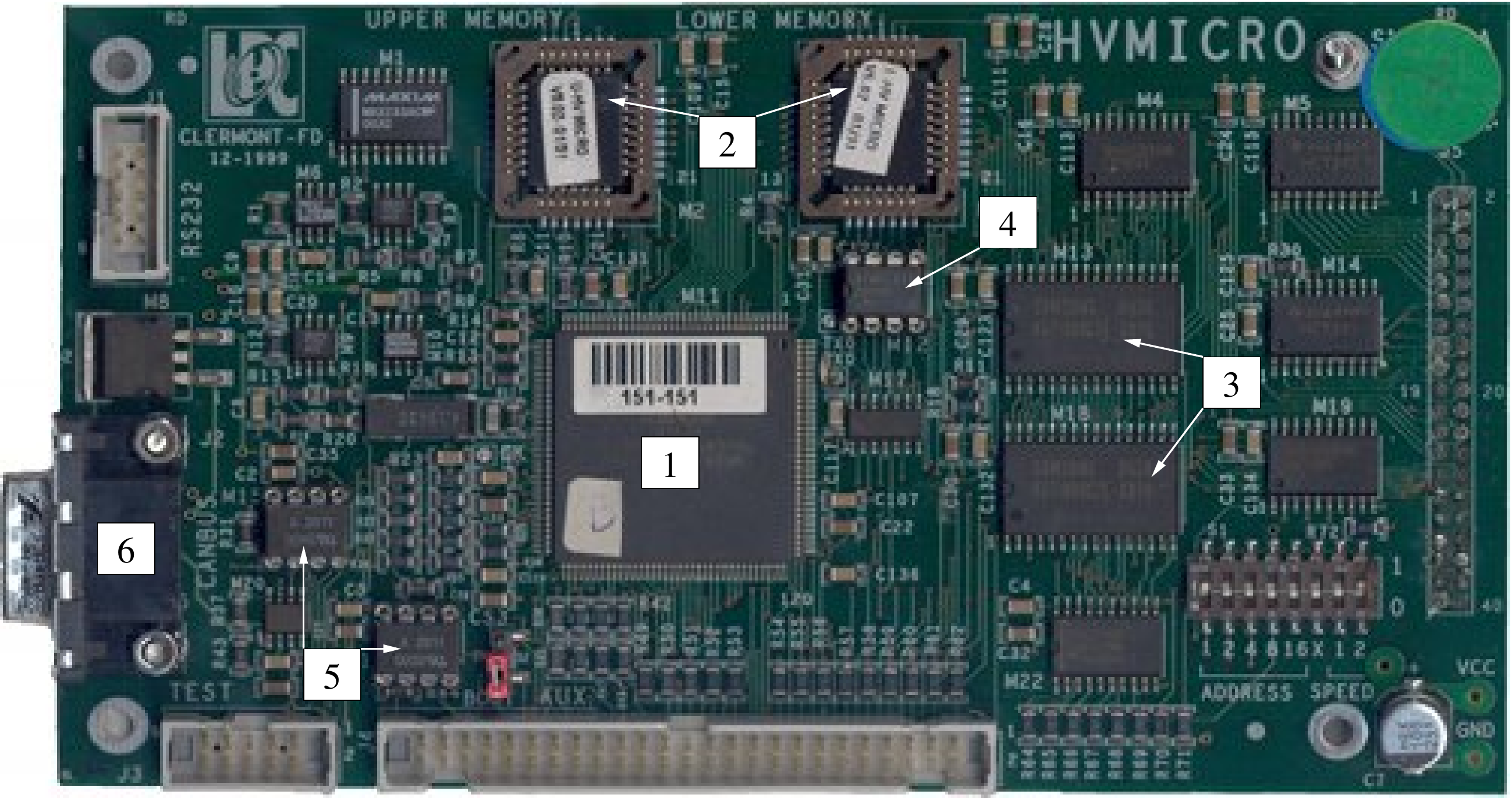}
\caption{Picture of the HVmicro board. The main components are: the 
micro-controller (1), the 256~KB flash memory (2), the 256~KB RAM (3), the 
2~KB EEPROM (4), the two opto-couplers of the CANbus interface (5) and the 
connector for the CANbus cable (6).}
\label{fig:HVmicro}       % Give a unique label
\end{figure}

The three externally generated input low voltages and the temperature probe located on the HVmicro board are monitored by a 10-bit ADC embedded in the micro-controller. The other values are monitored by ADCs located in the HVopto boards. 

For each monitored parameter, an average of 10 measurements is computed and stored in the HVmicro RAM (see Fig.~\ref{fig:scheme}). The comparison between this value and the nominal value allows determining the channel status (OK, warning or error) according to adjustable ranges for each channel.

The micro-controller program is stored in a 256~KB flash memory, while the 48 nominal HV values and the conversion parameters are stored in a 2~KB EEPROM. An additional 256~KB RAM contains all variables during the program execution and a copy of all the parameters of the EEPROM (see Fig.~\ref{fig:scheme}). In order to reduce SEE effects (see Section~\ref{sec:tolrad}), each parameter is copied five times in the RAM. Every time the program needs to access a parameter, the five copies are compared and, if they differ, the most frequent value is taken, the faulty values are corrected and a global status bit is set. The content of the EEPROM is verified prior to loading into the RAM, thanks to a checksum: in case of failure, default values are then used.

Finally, the CANbus interface integrated in the micro-controller is connected to the CANbus network through an opto-isolated interface. A picture of the HVmicro board can be seen in Fig.~\ref{fig:HVmicro}.

\subsection{HVopto board}
\label{subsec:HVopto}
Each HVopto board allows the regulation of the HV of 24 PMTs. 
%As shown in Fig.~\ref{fig:scheme} the input high voltage {\HVin} is first split in two, {\HVleft}  and {\HVright} powering the 12 left  and the 12 right PMTs respectively 
The input high voltage {\HVin} is distributed to the 12 left (\HVleft) and 12 right (\HVright) PMTs
(see Section~\ref{sec:Tile}). A  low voltage opto-coupler and a high voltage transistor allow to switch off each of the two HV values independently. In Fig.~\ref{fig:scheme}, they are shown as ``Odd switch'' and ``Even switch'' respectively. Regulation loops feed the 24 PMTs.  

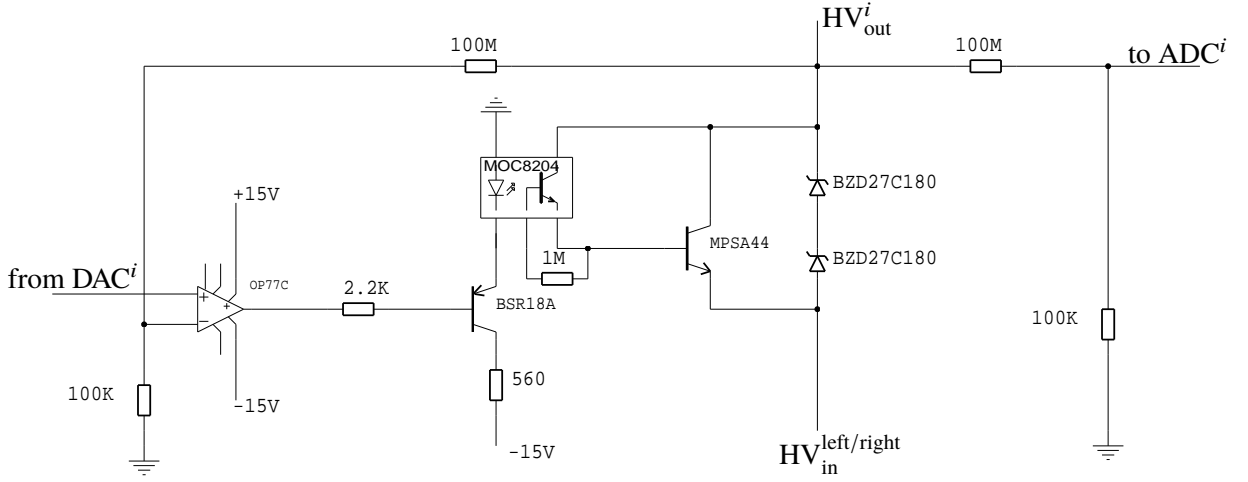
\begin{figure}[t]
% Use the relevant command for your figure-insertion program
% to insert the figure file.
  \centering
  \setlength{\unitlength}{0.1\textwidth}
  \begin{picture}(10,4.4)
    \put(0.05,0){\includegraphics[width=\textwidth]{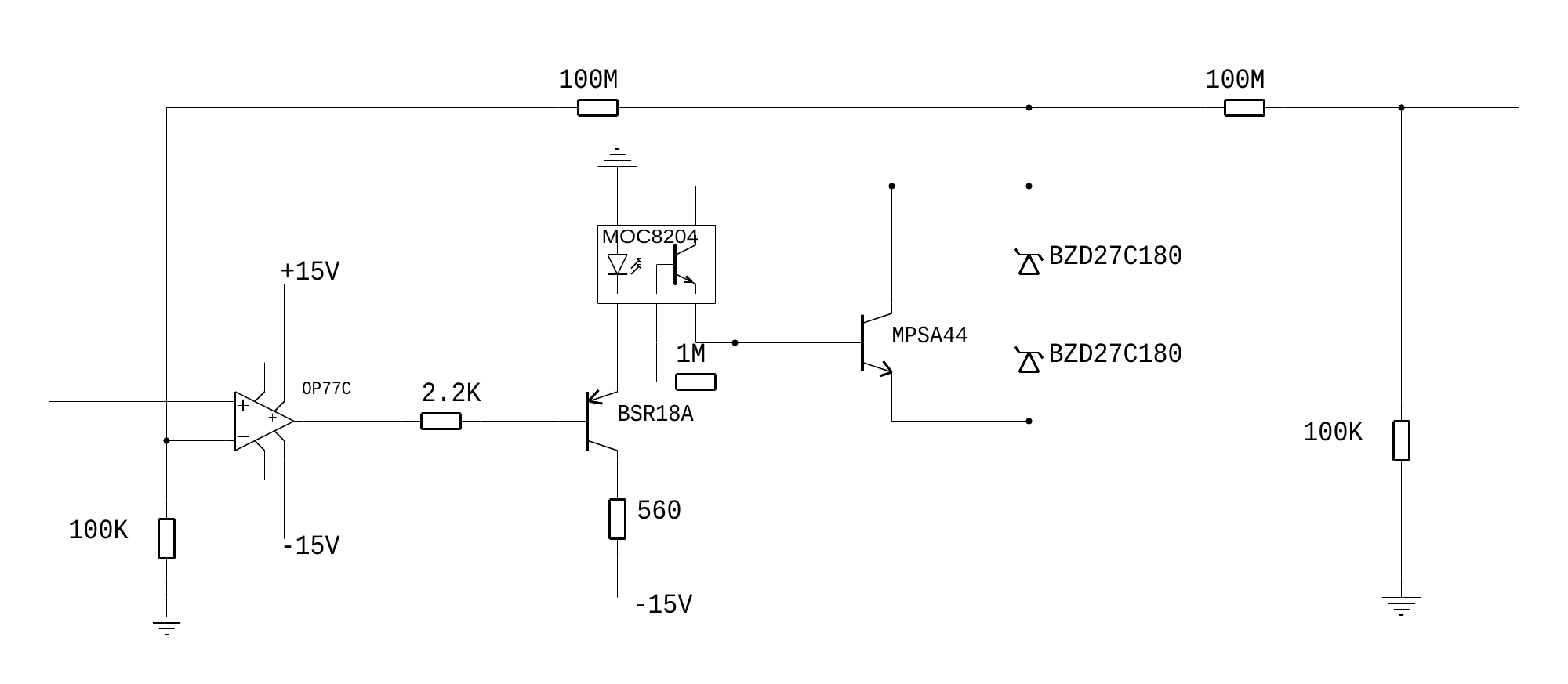}}
%    \multiput(0,0)(0,1){4}{\line(1,0){10}}
%    \multiput(0,0)(1,0){10}{\line(0,1){4}}
    \put(6.65,4){\HVout}
    \put(6.3,0.4){\HVleftright}
    \put(0,1.85){from \DAC}
    \put(9.15,3.7){to \ADC}
  \end{picture}  
  \caption{Scheme of the HVopto regulation loop: the %{\HVin}
    {\HVleftright}, coming from the HVPS, is modified by the opto-coupler (MOC8204) according to the DAC value, in order to have an {\HVout} value (going to the PMT) equal to the required one. This output value is monitored by the ADC at the right side of the scheme.}

  % \caption{Electronic scheme of the HVopto board regulation loop. The right hand side shows the monitoring of the output HV.}
  \label{fig:regulation}       % Give a unique label
\end{figure}
A HVopto board contains 24 12-bit DACs (one for each regulation loop) powered by the external $+15$~V and a locally generated $-5$~V. The negative voltage is provided by two regulators, each of them powering a group 12 DACs. In each regulation loop (see Fig.~\ref{fig:regulation}) a  DAC (\DAC), corresponding to the required HV value  {\HVnom}, is set by the HVmicro. The conversion procedure applied in the micro-controller is described in Section~\ref{subsec:conversion}. The DAC output is compared with a fraction of the output HV value,  {\HVout}, using an operational amplifier. The output current of this op-amp (OP77C) is then used (via the BSR18A transistor) to drive the LED of an opto-coupler (Motorola MOC8204) whose output transistor generates an increase of high voltage compared to the input negative value {\HVleftright}. A second transistor (MPSA44), creating a Darlington pair, allows the recovery of the gain of the opto-coupler. Finally, two Zener diodes (BZD27C180) with U$_Z = ($+$180 \pm 12)$~V are necessary to limit the high voltage at the bounds of the opto-coupler (maximal value of $+$400~V). With this design, the {\HVout} value applied to the PMT lies in the range:
\begin{equation}
%\mathrm{HV}_{in}+\mathrm{HV}_{max}^i > \mathrm{HV}_{out}^i > \mathrm{HV}_{in}+1~\mathrm{V}
{\HVleftright}+ {\HVmax} > {\HVout} > {\HVleftright}+1~\mathrm{V}
\label{eq:HVrange}
\end{equation}
where {\HVmax} is around $+$360~V with variations of about 20~V, due to the Zener diodes characteristics variations. The time recovery after a requested variation is less than 5 ms~\cite{HV_description}. In case the input low voltages are off, the applied voltage is {\HVout}=~{\HVleftright}+{\HVmax} and cannot be regulated anymore (reduced-HV mode). 

The applied value {\HVout} is converted by a 12-bit ADC through a voltage divider with a ratio of 1/1000, using a 100~M$\Omega$ HV resistor to a 12-bit number ({\ADC}, see Fig.~\ref{fig:regulation}) read by the micro-controller. This digital value {\ADC} is then converted to the monitored value {\HVmeas} (see Section~\ref{subsec:conversion}).

%monitored by a 12 bits ADC through a voltage divider with a ratio of 1/1000, using a 100~M$\Omega$ HV resistor.

\begin{figure}[t]
% Use the relevant command for your figure-insertion program
% to insert the figure file.
\centering
\includegraphics[width=\textwidth,clip]{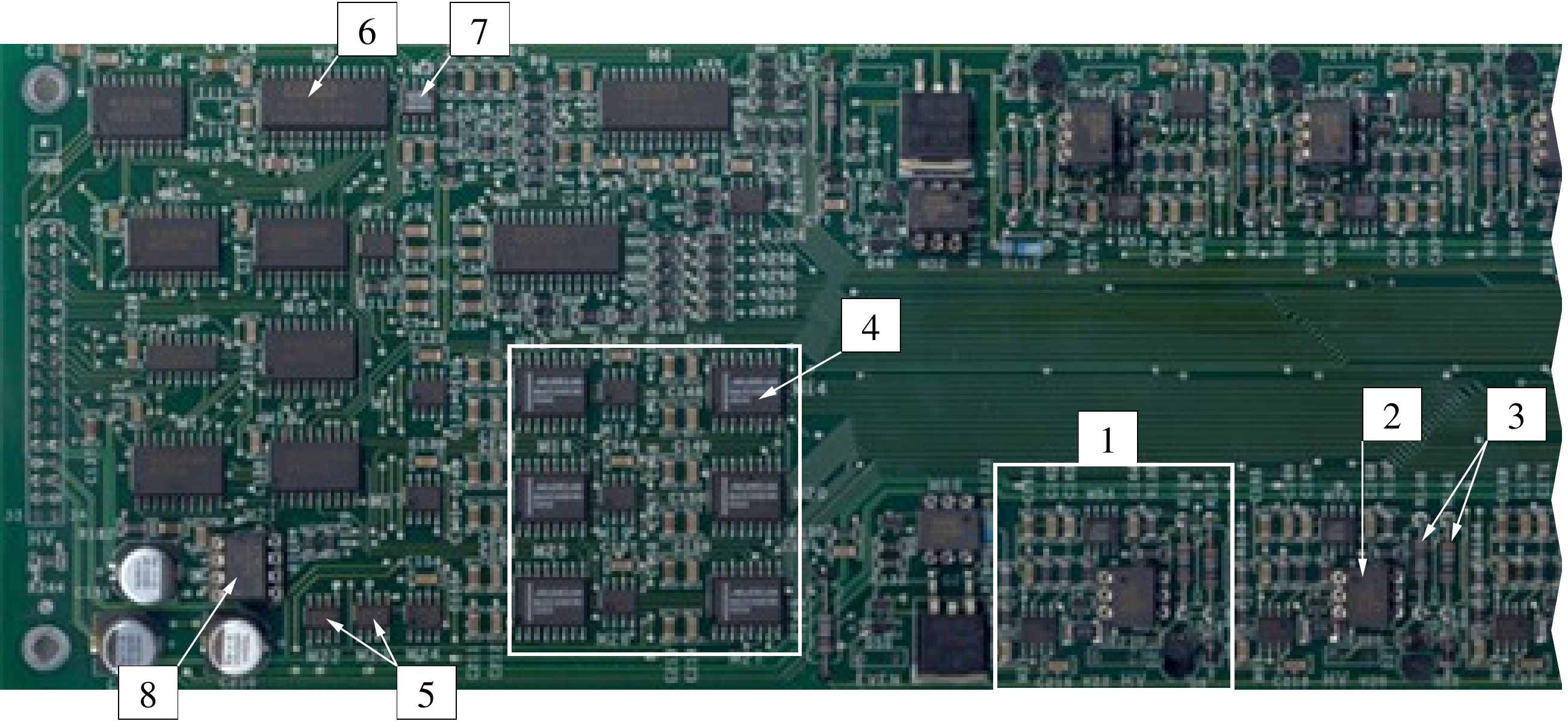}
\caption{Partial picture of the HVopto board showing four complete regulation loops. 
One of them is highlighted by the box (1).  One opto-coupler (2), the 
two 100~M$\Omega$ HV resistors (3), the six 
4-channels DACs highlighted by the box (4), the two $-5$~V regulators 
(5), the ADC (6), the voltage reference AD589 (7) and the 2~KB EEPROM (8) are also shown.}
\label{fig:HVopto}       % Give a unique label
\end{figure}

Each HVopto board contains a single ADC, multiplexed with 32 channels in order to monitor the mentioned 24 {\HVmeas}$'$s, the {\HVleft}  and {\HVright}, the two $-5$~V regulator outputs and the three temperature probes.
The reference voltage of the ADC (REF43) has a precision of 0.10\%.
In order to monitor the stability of the HV monitoring ADC (including its multiplexer and amplifier), a 1.23~V voltage reference AD589, is present on each HVopto board (see Fig.~\ref{fig:scheme}). The same digitization chain monitors this reference voltage and the voltages applied to the PMTs. Therefore, any drift of the measurement chain, in particular due to radiations, should translate into a drift of the monitored reference voltage. Such a drift has never been detected until now during the ATLAS operation (see also Section~\ref{sec:HVstability}). The board contains also a 2~KB EEPROM, that stores the conversion parameters of the DACs and ADC, to be retrieved by the micro-controller. A picture of the HVopto board can be seen in Figure~\ref{fig:HVopto}.

\subsection{Conversion of the DAC and ADC values}
\label{subsec:conversion}
The HVmicro board converts the required voltage values into integer values to be set in the DACs and integer ADC values into voltages or temperatures. Due to variations in the characteristics of the electronic components, the parameters of these conversions are different for each DAC or ADC and must be calibrated. The calibration parameters are stored in EEPROMs located on the relevant board (HVmicro or HVopto). All the boards have been calibrated once, during their production. Boards taken out of the detector and sent for maintenance are re-calibrated.

%The HVmicro board must perform conversions between the physical parameters, voltages and temperatures, and the digital values provided by the ADCs or to be set in the DACs. For the low voltages and temperatures conversions are linear and two parameters are needed: the slope and the offset. Since the characteristics of the temperature probes may suffer from variations, a calibration procedure must be applied and the two resulting parameters, obtained for each of them, are stored in the EEPROMs.
 %In the case of the high voltages the conversions are not linear and a second order polynomial is needed in order to improve the precision of the system.  

%, the non-linearity of the conversion must be compensated, in order to improve the precision of the system. For simplicity, the non-linearity is modelled by a quadratic term, therefore a second order polynomial is needed. The corresponding three parameters are stored in the HVopto board EEPROMs.
%the conversion is not linear because of the opto-coupler in the regulation loop. A second order polynomial is then used to linearize. The corresponding three parameters are stored in the HVmicro board EEPROMs. 

For each DAC channel, the calibration is performed by setting increasing values {\DAC} on the regulation loop.  The corresponding output HV values are determined making use of a  voltmeter (\HVout) and the monitoring ADC (\ADC). A second order polynomial $P_2$({\DAC}) is adjusted to the measured points ({\DAC}, {\HVout}). The resulting parameters are used by the HVmicro to convert the required high voltage value {\HVnom} to be applied to the PMT  into an integer value to set the corresponding DAC. The DAC precision allows a precision on {\HVnom} of 0.2 V.

A fit of a second order polynomial function on the measured points ({\HVout}, {\ADC}) is also used to extract the parameters needed by the HVmicro to convert the output of the ADCs ({\ADC}) into the measured HV  value ({\HVmeas}). 
The accuracy of the HV measurements is between 0.05 V and 0.1 V.

\begin{figure}[t]
  \begin{center}
    \includegraphics[width=0.495\textwidth]{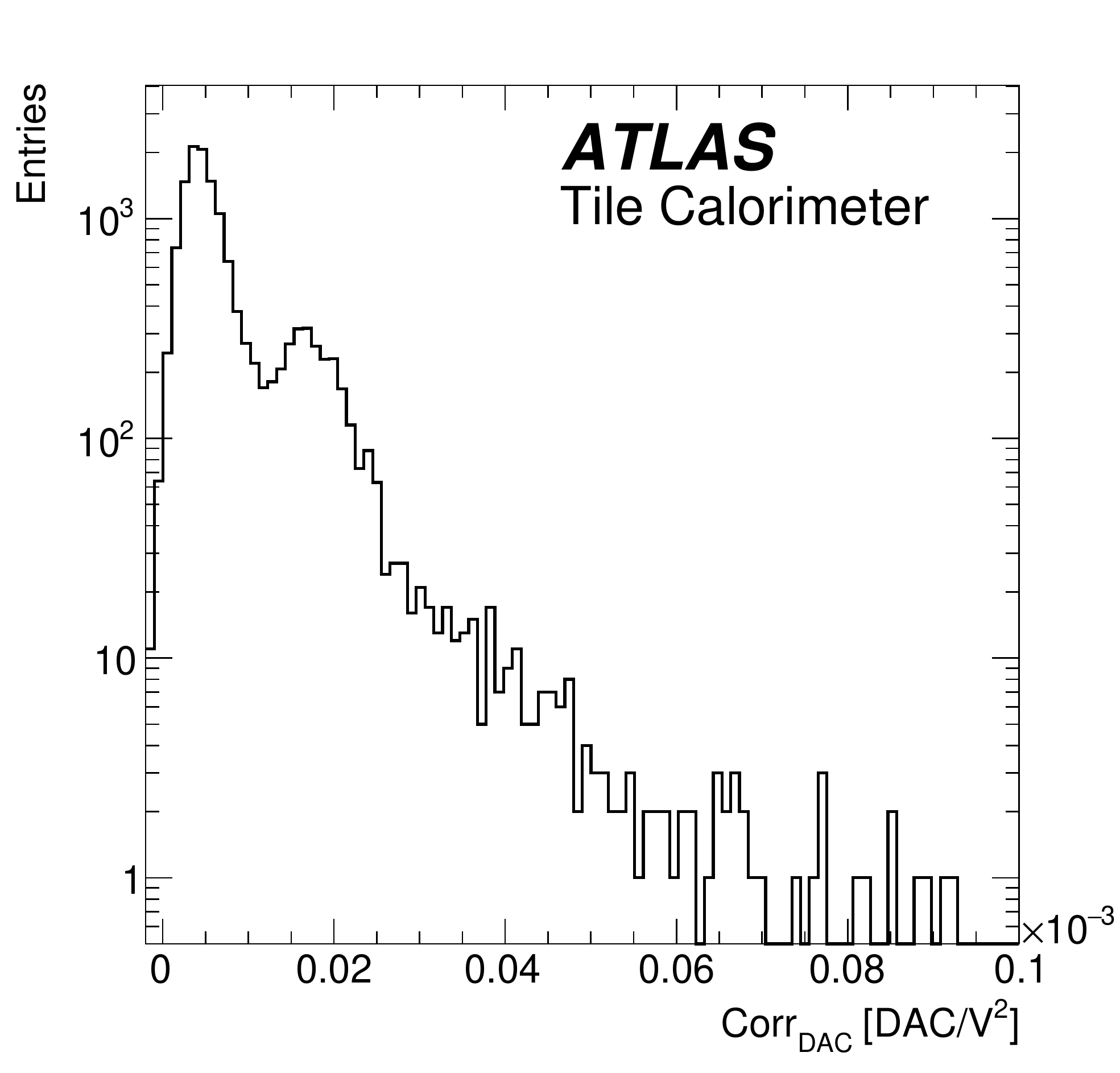}
    \includegraphics[width=0.495\textwidth]{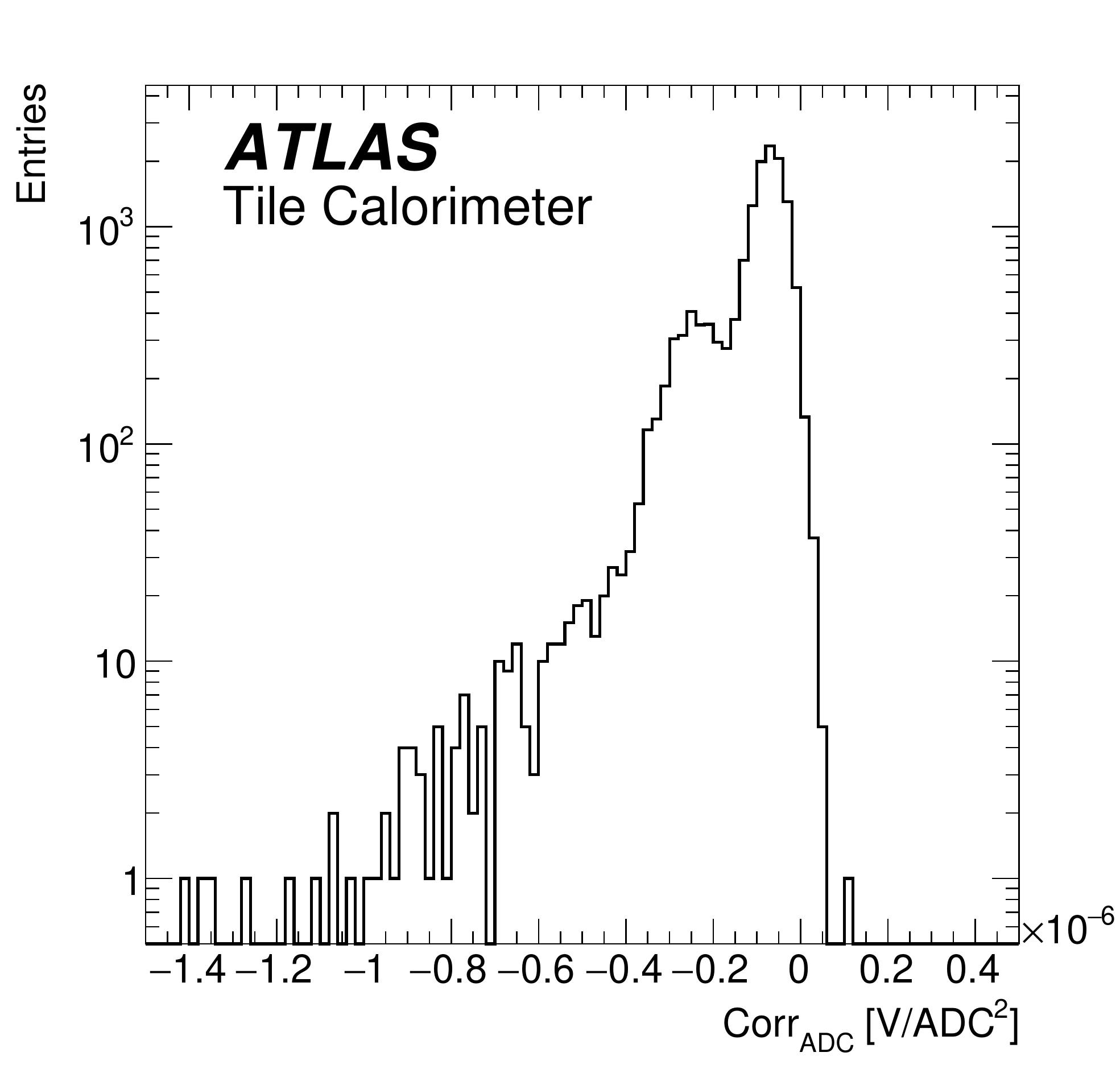}
    \caption{Distributions of the values of the quadratic term of the polynomial function used for conversions of the DACs (left) and ADCs (right) of all the HVopto boards.}\label{fig:ADCDACcor}
  \end{center}
\end{figure}

%A similar procedure is applied for the temperature measurements. 
% and slightly smaller than 0.1$^\circ$C for the voltage and temperature measurements respectively.
Figure~\ref{fig:ADCDACcor} shows the distributions of the values of the quadratic term of the polynomial function for the DAC and ADC conversions of all HVopto boards. While most of the channels exhibit a small correction, hence a behaviour close to be linear, a non negligible fraction
%, especially the DAC conversions, 
requires this second order correction.

The conversions between low voltages and temperatures are linear and two parameters are needed: the slope and the offset.
%A calibration of the temperature probes is also performed and
The precision on the temperature measurements is around 0.1$^\circ$C.

%\begin{figure}[ht13
% \begin{center}
 %\includegraphics[width=\textwidth]{CalibDAC.pdf}
 %\caption{Example of calibration of a DAC. The upper plot shows the DAC value as a function of the measured HV (crosses), together with a linear fit. The bottom plot shows, as a function of the target HV value, the difference between the correct DAC value, from the upper plot, and the computed DAC value, either from the linear fit (black circles) or the polynomial fit (white squares).}\label{fig:CalibDAC}
%\end{center}
%\end{figure}

%\begin{figure}[ht]
%\begin{center}
%   \includegraphics[width=\textwidth]{CalibADC.pdf}
%\caption{Example of calibration of an ADC. The upper plot shows the output HV as a function of the measured ADC value (crosses), together with a linear fit. The bottom plot shows, as a function of the measured ADC value, the difference between the real output HV, from the upper plot, and the computed HV value, either from the linear fit (black circles) or the polynomial fit (white squares).}\label{fig:CalibADC}
%\end{center}
%\end{figure}

\subsection{HVbus board}
\label{subsec:HVbus}
\begin{figure}[t]
% Use the relevant command for your figure-insertion program
% to insert the figure file.
\centering
\includegraphics[width=\textwidth,clip]{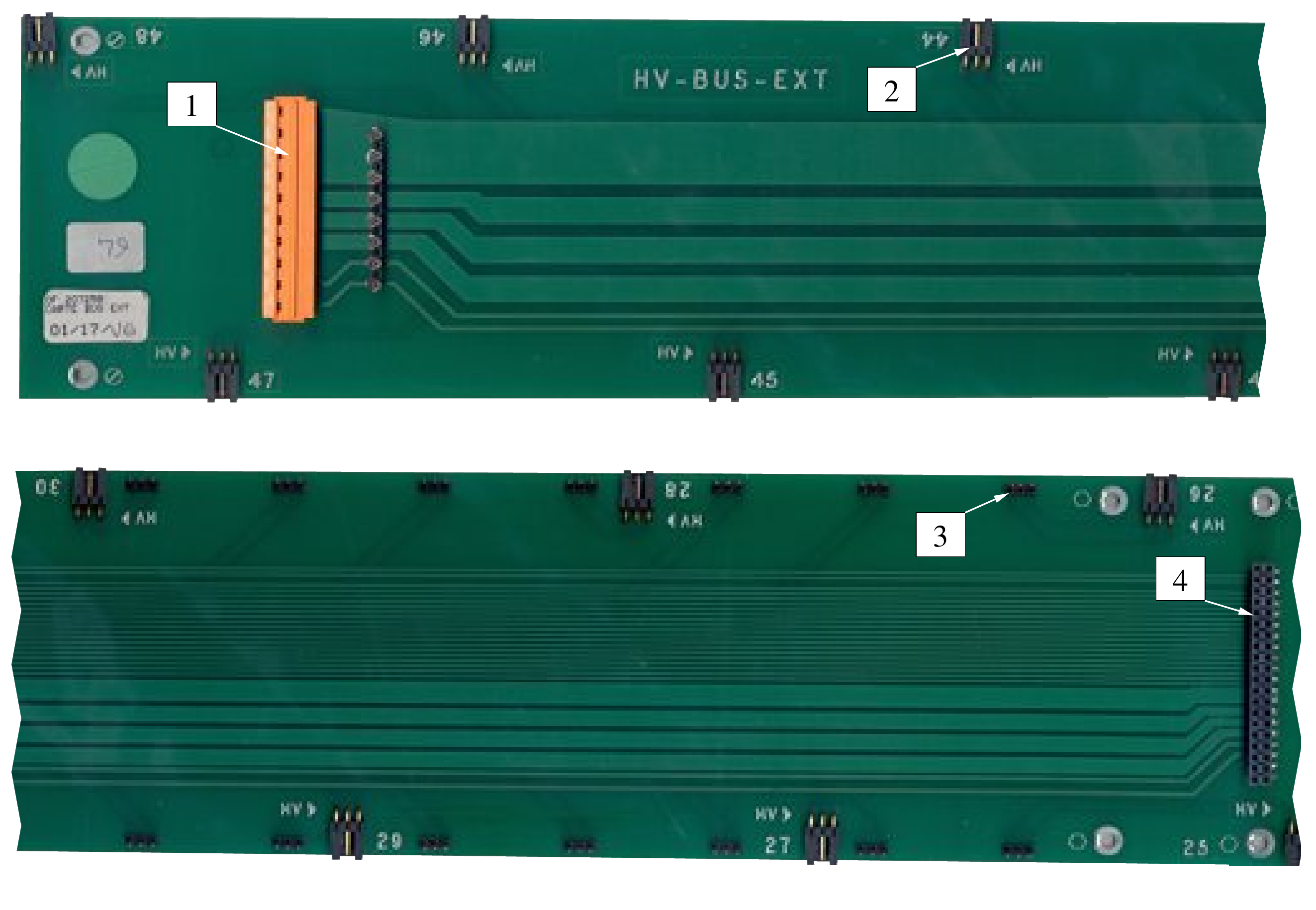}
\caption{Two partial views of the external HVbus board. In the top one can see the low and high voltage 
input connector (1) and several connectors feeding the HV to the PMTs, among them the one of PMT 44 is indicated with (2). The picture in the bottom shows several HV connectors linking the HVopto board to the 
HVbus board (for example the one indicated with (3)). The connector (4) links the external HVbus board with 
the internal HVbus board through the HVflex (see Fig.~\ref{fig:diagram}) that does not appear in the picture.}
\label{fig:HVbus}       % Give a unique label
\end{figure}

\begin{figure}[t]
\centering
\includegraphics[width=0.5\textwidth,clip]{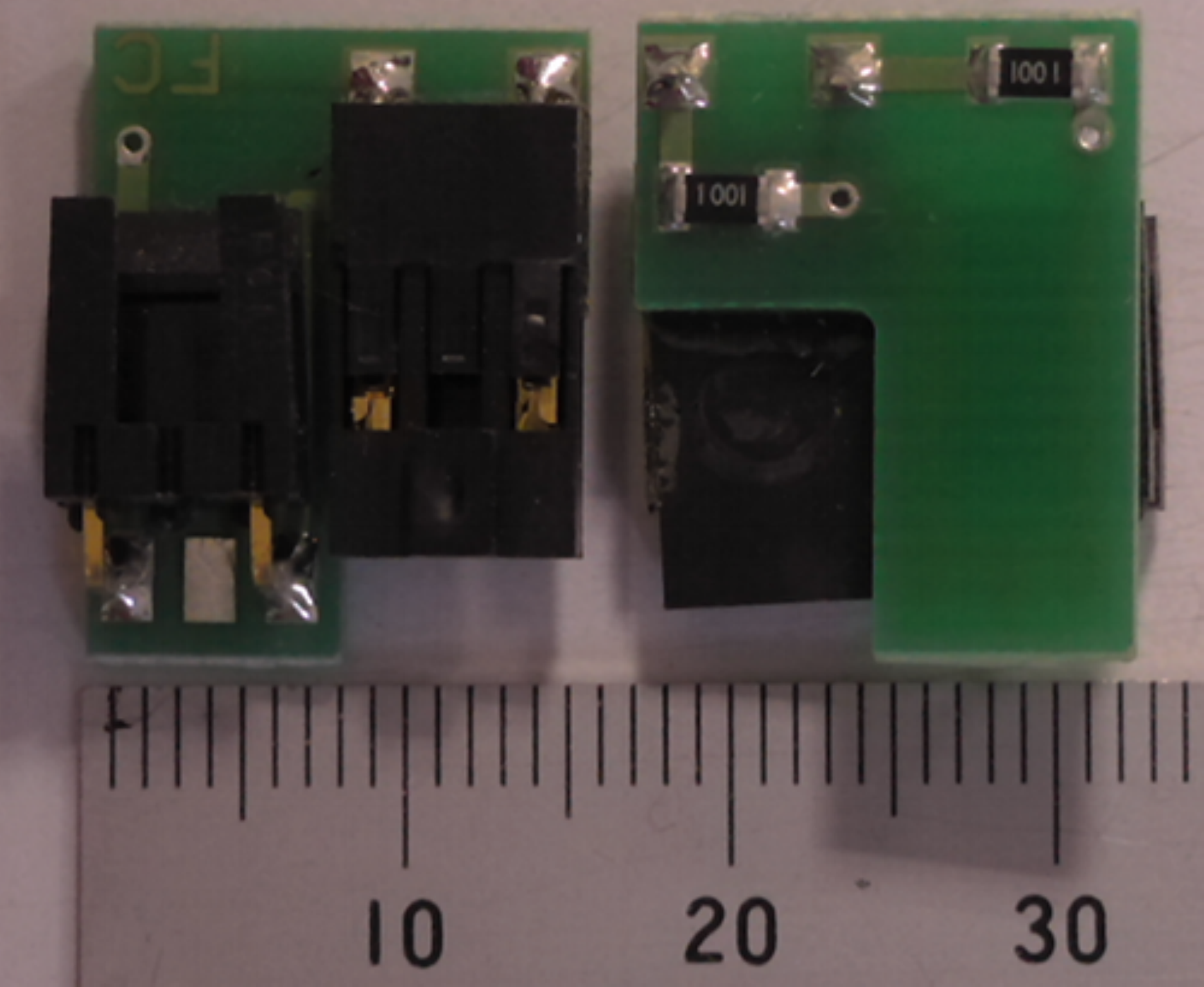}
\caption{The noise filter inserted in the connectors between the HVopto board and the input of each PMT voltage divider.}
\label{fig:NoiseKiller}       % Give a unique label
\end{figure}

As already mentioned, the role of the HVbus boards, shown partially in Fig.~\ref{fig:HVbus}, is to support mechanically the HVmicro and HVopto boards and to interconnect these two boards together and with the PMTs. Each HVbus board contains one temperature probe. A passive filter made of 1 k$\Omega$ resistors is inserted on the power and on the return of the HVbus connector between the HVopto board and the input of each PMT voltage divider~\cite{tdr_tile}.  A picture of the passive filter is shown in Fig.~\ref{fig:NoiseKiller}.

\subsection{Power consumption}
\label{subsec:consuption}
The baseline power consumption of the low voltages is 19~W for a full super-drawer. When the \HVin{} is powered (at $-830$~V or $-950$~V), the power consumption of the HV power supply is about 0.3~W per channel. When the \HVin{} is 0, an extra consumption of 0.32~W per channel on the $-15$~V power supply is observed. Therefore, in the worst case, the total power consumption for a super-drawer is 34~W. The cooling system of the TileCal front-end electronics is described in~\cite{cooling}.

\subsection{Online control of the HV system}
\label{subsec:DCS}
All the HV systems of each of the four partitions are controlled and monitored independently by the DCS that was implemented with the PVSS application during LHC Run 1 (2009-2013). Since 2015 (LHC Run 2) it is using the WinCC OA application. Each partition is under the supervision of one computer. The communication with the HVPS is achieved using a commercial serial-to-Ethernet converter (in order to convert the ModBus/RTU protocol of the HVPS to the ModBus/TCP protocol) connected to the local TCP/IP network. The 64 super-drawers of the partition are grouped in four branches of 16 super-drawers. Within a branch, the super-drawers are daisy-chained on a single CANbus cable. The four CANbus cables were then connected to the four ports of a Kvaser PCIcan board during Run 1, replaced by a 16-channels SYSTEC USB/CANbus converter for Run 2. A dedicated C++ application implements the communication protocol with the HVmicro boards and acts as a server (DIM server for Run 1, OPC UA server for Run 2) while the DCS application is the client.

Monitoring and configuration services were implemented in the PVSS/WinCC OA framework. The monitoring service allows retrieving all the values from the HVmicro boards and storing them in a data base for offline analysis (see Section~\ref{sec:HVstability}). The configuration service allows setting the required HV values for each PMT as well as the allowed ranges of the monitored values. Warnings and alarms are generated when the values are outside the allowed range.

%-------------------------------------------------------------------------------
\section{Tolerance to radiation}
\label{sec:tolrad}
%-------------------------------------------------------------------------------

\begin{table}[t]
\centering
\caption{Required lower limit radiation tolerance values for the HV boards electronics components.}
\label{tab:radtol}       % Give a unique label
% For LaTeX tables you can use 
\begin{tabular}{	|l|l|c|} 
\hline 
 NIEL (1 MeV eq. n $\times$ cm$^{-2}$) & HVmicro &   3.3$\times 10^{11}$ \\ 
 \cline {2-3}
 & HVopto & 5.3$\times 10 ^{11}$  \\ \hline
 TID (Gy) & HVmicro &10.1  \\ 
  \cline {2-3}
  & HVopto & 26.9$$  \\ \hline
  SEE (>20 MeV p $\times$ cm$^{-2}$) & HVmicro & 8.6$\times 10 ^ {9}$  \\
  \cline {2-3}
 & HVopto & 3.4$\times10^{10}$  \\ \hline
\end{tabular}
%\end{center}
\end{table}
The location of the HV system in the detector implies that the choice of the electronics components must be performed according to the ATLAS Policy on Radiation Tolerant Electronics~\cite{PERF-2007-01}. The requirement is that the boards must be operational during 10 years at the nominal LHC Luminosity even if they are placed in the highest radiation level locations. The radiation values in the different regions of ATLAS were determined using  the ATLAS Monte Carlo radiation simulation~\cite{Dawson:1499645} program. Non Ionizing Energy Loss (NIEL), Total Ionizing Dose (TID) and Single Event Effect (SEE) were considered. The required lower values of the radiation intensity that the different components must bear in 10 years,  reported in Table~\ref{tab:radtol}, include large safety factors. 

Qualification tests aiming to determine the effects on the components of  NIEL, TID and SEE were performed using neutrons, gammas and protons beams respectively.   Neutrons with kinetic energy close to 1 MeV were used at the Prospero accelerator at Dijon in France~\cite{NIEL_tests}. The boards were also exposed to gammas with energy close to 1.2 MeV at Pagure in France~\cite{NIEL_tests} and to protons of 300 MeV at PSI in Switzerland~\cite{SEE_tests_1},~\cite{SEE_tests_2} and~\cite{SEE_tests_3}. 

As a consequence of the deterioration of the functioning of the HVopto boards during the SEE and NIEL tests, two major modifications of the original design were necessary~\cite{SEE_tests_1,NIEL_tests}: 
\begin{itemize}
\item
the original opto-coupler (MCT2) allowing switching on/off separately the HV of the two groups of 12 PMTs   was replaced by the opto-coupler MOC8204 and
\item
two transistors were placed at each side of the opto-coupler MOC8204 
in the regulation loop (see Fig.~\ref{fig:regulation}) 
to balance the gain losses of the optic and of the electronic sides of the opto-coupler.
\end{itemize}
Other components were affected by radiations, but without spoiling the performance of the system. In particular a drift on the ADC signal of 1-2\% was observed during the TID tests. This effect can be easily detected, monitoring the 1.23~V voltage reference AD589, as explained in Section~\ref{subsec:HVopto}.

The tolerance to radiation of the whole system was found to be at least 100 Gy to gammas, 1.8 $\times~10^{12}$~MeV eq. n $\times$ cm$^{-2}$ to low-energy neutrons and 1.8 $\times10^{11}$MeV eq. n $\times$ cm$^{-2}$ to high-energy neutrons. The described implementations made all the components safe against SEEs as well. The measured tolerance lower limits are bigger than the required upper limits.

%-------------------------------------------------------------------------------
\section{Characteristic of the HV system}
\label{sec:characteristic}
%-------------------------------------------------------------------------------
\subsection{Sensitivity to Low Voltage variations}
\label{subsec:sens-LV}
Using the DCS information (Secion~\ref{subsec:DCS}), systematic studies were performed to check the stability of the HV system  response when the effective low voltages values supplied by the LV(+5~V) and LV($\pm$ 15~V) sources vary. The effects on the temperatures in the different points of a super-drawer, the values of {\HVin}  in each half-drawer,  the regulated {\HVout}, the reference voltage (1.23 V) in each drawer and the $-$5~V value of each drawer were investigated. Table~\ref{tab:sens-LV} shows the ranges of variation of the applied LV, $\Delta$LV(+5~V), $\Delta$LV(+15~V) and $\Delta$LV($-$15~V) that result in no variation of the considered quantities, within the precision of their measurements.
\begin{table}[t]
\caption{Ranges of variations of the effective applied Low Voltages (LV) resulting in stable values of the quantities in the first column. The values in the last column correspond to simultaneous variations of the voltages supplied by the two $\pm$ 15~V power supplies.}   
\label{tab:sens-LV}       % Give a unique label
% For LaTeX tables you can use 
\begin{tabular}{	|c|c|c|c|c|} 
\hline
& $\Delta$LV($-15$~V) [V]& $\Delta$LV($+15$~V) [V]& $\Delta$LV($+5$~V) [V]& $\Delta$LV($+15$~V) [V]$^{a)}$ and  \\ 
& & & &$\Delta$LV($-15$~V) [V]$^{b)}$ \\ \hline  \hline
Temperature probes  & [-17, -7]  & [9, 17]  & [4.4, 5.5] & $a)$ [7, 17], $b)$ [-17, -7] \\ \hline 
{\HVin}  in each drawer& [-17, -5] & [9, 17] & [4.7, 5.5] & $a)$ [5, 17], $b)$ [-17, -5]  \\ \hline 
{\HVout}   & [-17, -5]  & [11, 17]  & [4.4, 5.5] &  $a)$ [11, 17], $b)$ [-17, -11] \\ \hline  
Reference Voltage & [-17, -4] & [8, 17]  & [4.4, 5.5] & $a)$[5, 17], $b)$ [-17, -5] \\ 
  (1.23 V) &  &  & &   \\ \hline 
$-5$~V & [-17, -7]  & [8, 17]  & [4.4, 5.5] & $a)$ [7, 17], $b)$ [-17, -7]  \\ \hline  
\end{tabular}
%\end{center}
\end{table}

As expected, the acceptable voltage ranges are larger for the $\pm$15~V power supplies. Absolute values larger than 17 V affect the safety functioning of some components  as the DACs connected to the regulation loops. For the +5~V power supply, which is mainly used in the logic part of the boards, values smaller than 4.4 V affect the micro-controller that hosts the CANbus interface. These results allow defining the DCS warning and alarm ranges used to monitor online the low voltage values during data taking and making diagnosis of the observed failures.

\subsection{Sensitivity to High Voltage variation}
\label{subsec:sens-HV}
The sensitivity of {\HVout} to the {\HVin} value was evaluated, by varrying the value of {\HVin} in the allowed range $[\HVnom+\HVmax,\HVnom+1~\mathrm{V}]$ (see Eq.~(\ref{eq:HVrange})), and simultaneously measuring {\HVout} with a multi-meter. Since no change of {\HVout} was measured, one can conclude that the sensitivity is smaller than the accuracy of the multi-meter, i.e. 0.1~V.
%When the high voltage power supply value is changed within the allowed regulation range (see Eq.~(\ref{eq:HVrange})), multi-meter measurements show that there is no effect on the output {\HVout}. One concludes that the sensitivity to the input HV in the considered variation range is less than the accuracy of the used multi-meter, 0.1~V.

\subsection{Sensitivity to temperature variation}
\label{subsec:sens-temp}
A temperature-regulated box has been used to measure the sensitivity of the HV distribution system to the temperature in the drawers~\cite{HV_description}. During the tests, the temperature of 12 PMTs was measured with three silicon probes with an accuracy of 0.1$^\circ$C. The temperature of the box  was varied in the range 20-35$^\circ$C with steps of 5$^\circ$C. For each temperature value the measurements lasted 10 hours. A linear behaviour of the average values of the HV measurements as a function of the temperature was observed. The larger  slope is $+$0.03~V/$^\circ$C, much smaller than the specification value of Section~\ref{sec:HV} (0.2~V/$^\circ$C). 
%During the ATLAS data taking period spreads of about 0.1$^\circ$C were observed.
%The obtained sensitivities (the slope of the linear behaviour) of the HVs  at the input of the DAC,  $HV_{in}$, at the output of the regulation loop, $HV^i_{out}$, and at the output of the ADC, $HV^i$, are reported in Table~\ref{tab:sens-temp}. These variations are very small with respect to the specification of Section~\ref{sec:HV} (< 0.2/V$^\circ$C). The distributions of the measured temperatures in the super-drawers during Run 1 show a spread of 0.1$^\circ$C. 
%\begin{table}[h]
%\centering
%\caption{Sensitivity of the HV measured value to the temperature.}
%\label{tab:sens-temp}       % Give a unique label
% For LaTeX tables you can use 
%\begin{tabular}{	|c|c|c|} 
%\hline
%$HV_{in}$ & $HV^i_{out}$ & $HV^i$\\ \hline  \hline
%-0.011 V/$^\circ$C  & +0.028 V/$^\circ$C & -0.014 V$^\circ$C \\ \hline 
% \end{tabular}
%\end{center}
%\end{table}

\subsection{Sensitivity to humidity variation}
\label{subsubsec:humidity} 
 Sensitivity of the HV system to the environment humidity has been studied at CERN. At the test bench all the elements of a full super-drawer, apart  the HV and LV power supplies, were placed inside a climate chamber. For low-level humidity (below 60\%), no HV trip occurred on the 48 channels over a period of 8 weeks. Several trips occurred after 7 (24) hours in the case of humidity values above 70\% (between 60 to 70\%). On the base of these results, it was decided to inject dry air in the girder during the ATLAS operations to maintain the humidity below 60\% (see Section~\ref{sec:HV}). 

\subsection{HV ripple at the divider output}
\label{subsec:ripple}
%The HV noise at the PMT divider to which the HV is applied is 1.40 $\pm$ 0.10 mV~\cite{1748-0221-11-02-C02050} . It corresponds to the RMS of the noise distribution of 12 dividers obtained by using a 1.5 m long multi-conductor HV cable at the place of the HVopto board that has a shorter HV connector. The noise value corresponds to a HV ripple lower than 9.70 $\pm$ 0.70 mV, within the specification of 20 mV (see Section~\ref{sec:HV}).
The ripple of {\HVout} at the level of the PMT divider has been measured with an oscilloscope in a slightly different condition than the standard configuration, since the cable between the divider and the HVopto board was a multi-conductor cable 1.5~m long instead of the HVbus board and a few centimeters of HV cable. The measured ripple is 9.70$\pm$0.70~mV, within the specification of 20~mV (see Section~\ref{sec:HV}).

%\section{Analysis of faulty boards}

%-------------------------------------------------------------------------------
\section{Analysis of faulty boards after Run 1}
\label{sec:faulty}
%-------------------------------------------------------------------------------

%At the end of the LHC run 1, a 2-year period of maintenance was possible (Long Shutdown 1, LS1) during which all super-drawers were taken out of the calorimeter to be repaired and tested.

%In total during LS1, 104 HV boards were repaired. The most frequent defect was a drift of the resistance value of a 100~M$\Omega$ high voltage resistor (116 occurences), with 90\% of them with a variation smaller than 1~\%. The two HV resistors (from the regulation loop and the reading voltage divider, see Figure~\ref{fig:regulation}) are equally failing. Only eight opto-couplers failed and nine output transistors were destroyed, the latter due to short circuits in the voltage divider board of the PMTs.

At the end of the LHC Run 1, February 2013, a 2-year period of maintenance took place.  During the Long Shutdown 1 (LS1) all the super-drawers were removed from the calorimeter to be tested and repaired if necessary.

A total of 104 HV boards did not pass the quality assessment and were repaired. The most frequent defect was a drift of the resistance value of a 100~M$\Omega$ high voltage resistor  of the HVopto board (116 occurrences). The two HV resistors (from the regulation loop and the reading voltage divider, see Fig.~\ref{fig:regulation}) were equally failing. A large fraction of these faulty resistors, 90\%, shows a variation smaller than 1\%.  Only eight opto-couplers failed and nine output transistors were destroyed due to short circuits in the voltage divider board of the PMTs.

%\begin{table}[h]
%\centering
%\caption{Summary of the  fails in the 161 replaced HV boards during Run 1.}
%\label{tab-6}       % Give a unique label
% For LaTeX tables you can use 
%\begin{tabular}{	|c|c|} 
%\hline
%Sources & Failed boards  \\ \hline  \hline
%HV resistance  & 116\\ \hline 
%Opto-coupler  & 20\\ \hline
%Output transistor  & 9\\ \hline 
%Connector  & 8\\ \hline
%EEprom  & 3\\ \hline 
%Shorts  & 2\\ \hline
%DAC  & 2\\ \hline 
%Dirtiness  & 8\\ \hline
%\end{tabular}
%\end{center}
%\end{table}
% Use the relevant command for your figure-insertion program
% to insert the figure file.
%\centering
%\includegraphics[width=7cm,clip]{fig_10left.pdf}
%\includegraphics[width=7cm,clip]{fig_10right.pdf}
%\caption{(left) Resolution of the $E^\mathrm{miss}_{x(y)}$ determinations as a function of the number of primary vertices. (right) Mean value of the projection of $E^\mathrm{miss}_\mathrm{T}$ along the $Z$ particle transverse momentum, $p^Z_\mathrm{T}$, as a function $p^Z_\mathrm{T}$. A bin is required to have a minimum of 200 events to be considered for the plots~\cite{Ref3}.}
%\label{fig-10}       % Give a unique label
%\end{figure}
%\subsection{The $E^{miss}_T$ measurement scale}
%\label{subsec-4.4}

\section{HV stability during 2015 and 2016 ATLAS run at LHC}
\label{sec:HVstability}

The HV stability is the most important feature of the system because it affects the stability of the PMTs gains. The relation between the PMT gain and the applied high voltage is given by
\begin{equation}
\Gi= \alpha^i \times ( {\HVout} )^{\beta^i} 
\label{eq:G}                                       
\end{equation}
where $\alpha^i$ and $\beta^i$ are parameters specific to each PMT. In TileCal $\beta^i$ $\approx$ 7. A variation of the high voltage induces a gain variation deteriorating the measurement of the energy deposited by the particles in the calorimeter cells. 
The HV stability has been studied using HV dedicated test benches, at test-beam~\cite{HV_2001} and in ATLAS. 

%\subsection{HV stability monitoring at Beam Tests}
%\label{subsec:stabilityTB}

%The results obtained analyzing  data collected during the 1997, 1998 and 2001 beam tests are presented here. During the 1997 (1998) data taking 56 (90) PMTs were monitored for a period of about 6 months.  The HV fluctuations of the {\ADC} measurements are much smaller than the specification value of Section~\ref{sec:HV} (0.5 V): $\sigma_{{\ADC} }$ = (0.01 to 0.11) V  and $\sigma_{{\ADC} }$ = (0.05 to 0.13) V~\cite{HV_description}. 
%The results obtained at the 2001 Test Beam at CERN can be summarized as follow~\cite{HV_2001}: a) differences between HV orders and read values are on average of -0.1 V, corresponding to about 0.2\%, with a RMS of 0.7 V, b) concerning the HV stability, the RMS of the PMTs distributions were characterized by a RMS smaller then 0.14 V, well inside the specifications.
%They correspond to a follow up of the HV over about 6 months, comparable to two years of ATLAS data taking at LHC.

%-------------------------------------------------------------------------------
%\subsection{HV stability monitoring during 2015 and 2016 ATLAS run at LHC}
%\label{subsec:stabilityATLAS}
%-------------------------------------------------------------------------------  
\subsection{Identification of unstable channels}
\label{subsec:unstable}
\begin{figure}[t]
  \begin{center}
    \includegraphics[width=0.495\textwidth]{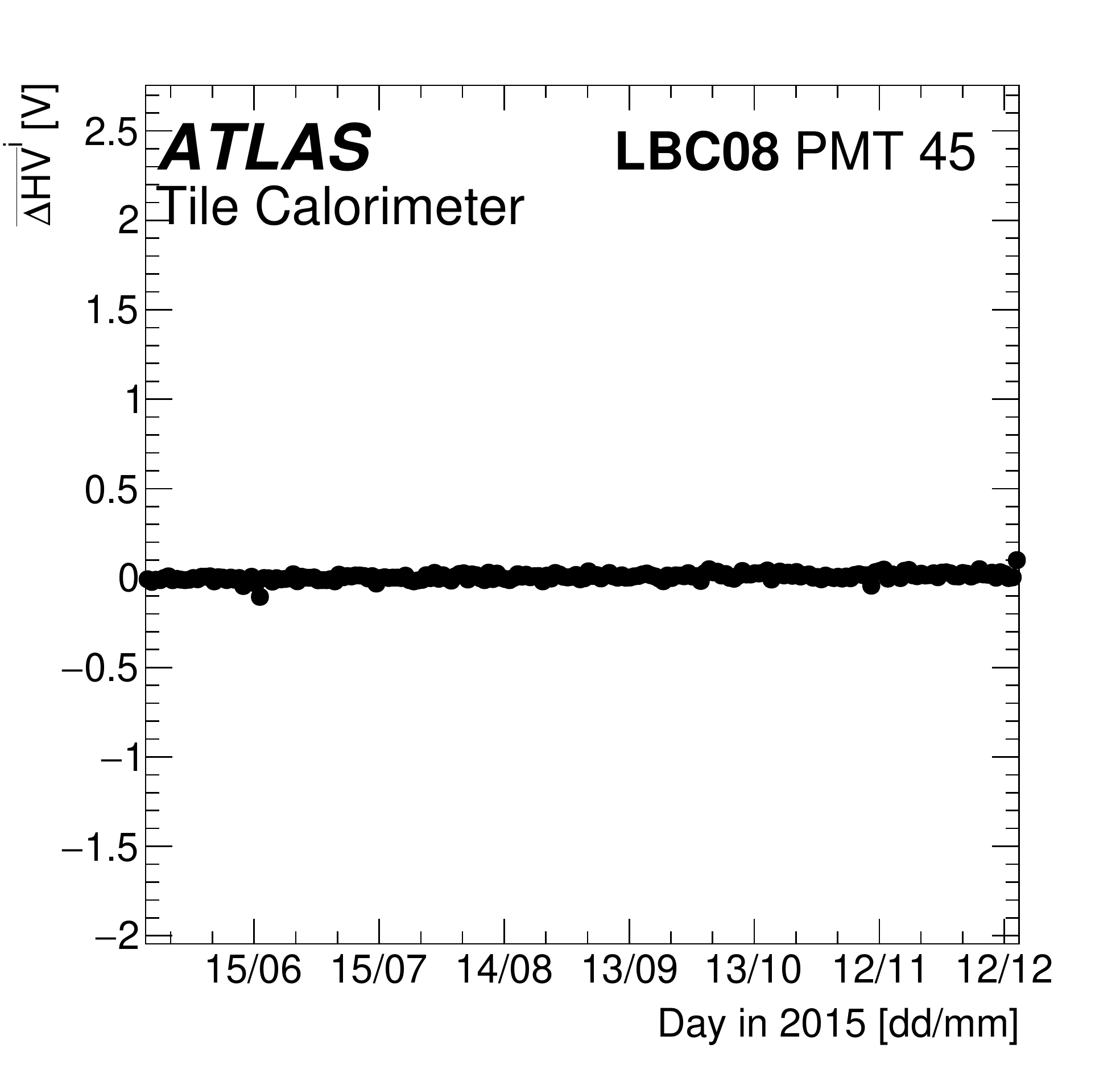}
    \includegraphics[width=0.495\textwidth]{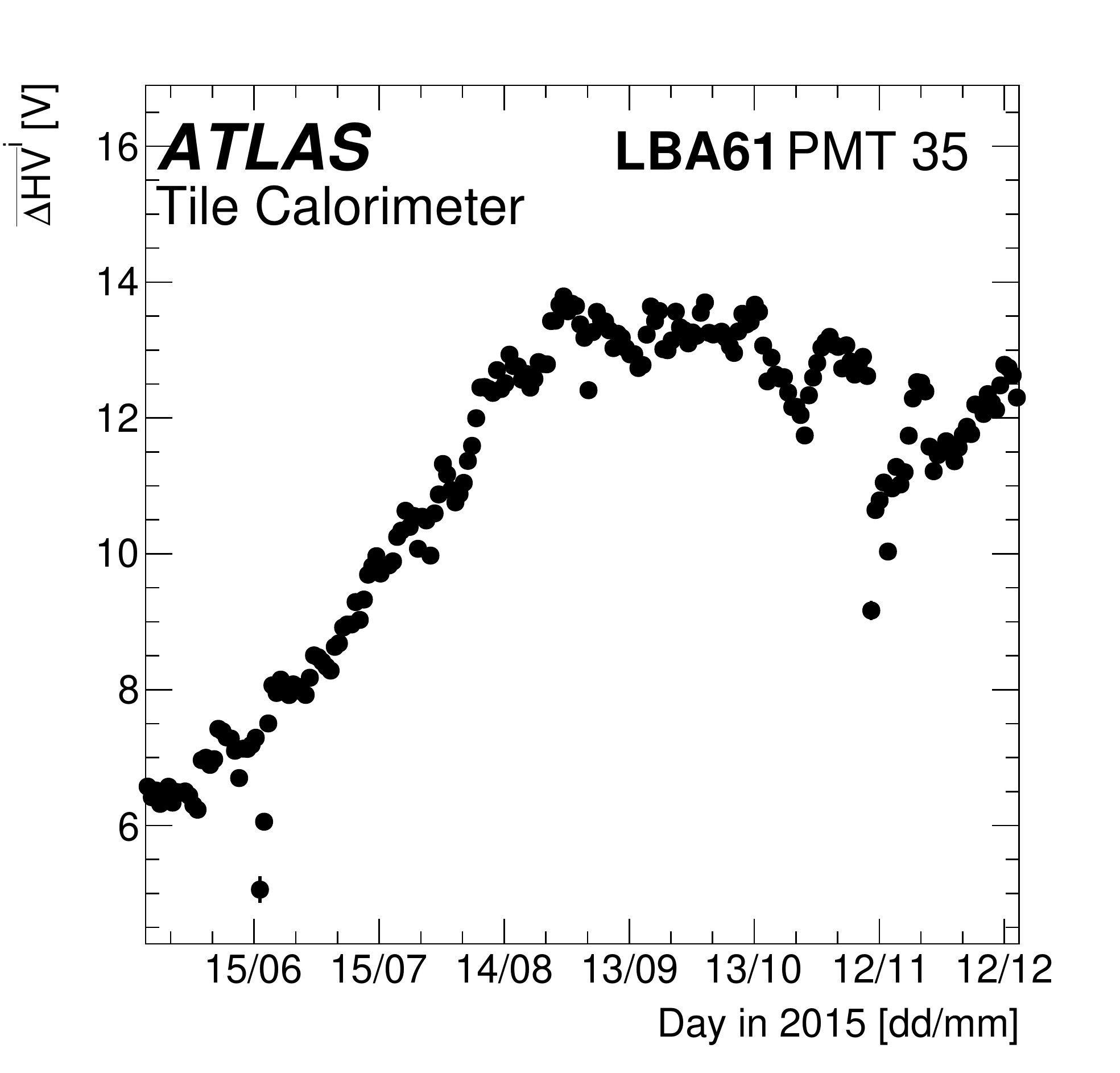}
    \caption{Behaviour of \AvDeltaHV{} as a function of time of a stable (left) and an unstable channel (right) observed during the 2015 run.}\label{fig:stableunstable}
  \end{center}
\end{figure}

During data taking periods, the DCS monitors the HV values that are applied to the PMTs by communicating with the HVmicro boards using the CANbus protocol. Alarms are automatically generated to alert the shifter in case the applied HV value drifts outside the predefined tolerance range. The monitored values are also stored in a database, allowing later offline analysis. However, in order to save storage space, only values that differ more than 0.5 V from the previous measurement are recorded. At least one measured value is stored every hour. Results of the analysis of 2012 data can be found in~\cite{Valery:1743602}. Below results of the analysis  of data collected from May 5th 2015 to December 15th 2015 and from April 19th 2016 to December 6th 2016 are discussed. The number of channels that could be monitored in 2015 (2016) was 9593 (9613).
Only few channels show a behaviour that does not correspond to the requirements discussed in Section~\ref{sec:HV}. These channels have been identified studying the differences 
\begin{equation}
\DeltaHV(t)= {\HVnom}(t) - {\HVmeas}(t)
\label{eq:offset}                                         
\end{equation}
%$\Delta HV = \mid HVi-HV^i_{out}\mid$
as a function of time $t$. The quantities {\HVnom} and {\HVmeas}  of each channel $i$ were defined in Section~\ref{subsec:conversion}. 
The \DeltaHV{} distributions were determined for 2015 and 2016 data respectively.  The most probable value, \mudhvi, and the spread, \sigdhvi, of the distributions were estimated as the parameters of Gaussian fits performed in the range  $\pm$ 0.5 V around the maximum of the distributions. In the case of very narrow distributions, ranging in a window smaller than 0.6 V, the values \mudhvi{} and \sigdhvi{} correspond to the mean and the RMS of the distributions respectively.
The unstable channels verify the condition 
\begin{equation}
%\mid  \mu_{part} -  \mu_i \mid \geq 5 \sigma_{part} 
\mid \AvDeltaHV-  \mudhvi \mid \geq 0.5\;\mathrm{V}
\label{eq:dif}                                         
\end{equation}
for more than 5\% of the days during the considered period of time. In Eq.~(\ref{eq:dif}) \AvDeltaHV{} is the average of the \DeltaHV{} measurements performed in a day. As an example, the behaviour of \AvDeltaHV{} as a function of time of a stable and an unstable channel observed during the 2015 run are shown in Fig.~\ref{fig:stableunstable}. 
A total of 24 (18) unstable channels were identified in 2015 (2016). 

In order to detect channels that would have a stable \AvDeltaHV{} but with short term instabilities, the \sigdhvi{} of each stable channel must be at most 0.5~V and less than 20\% of the \DeltaHV{} values are allowed outside the range $\mudhvi\pm10\sigdhvi$. All channels that fail these conditions are also detected as unstable by condition~(\ref{eq:dif}).

Among the unstable channels, 11 were unstable both in 2015 and 2016. A total of 13 channels  were unstable in 2015 but not in 2016. The exact reason why these channels recovered in 2016 is not understood except for three channels:  two are subject to sudden jumps of \HVout{} (probably due to a failure of the DAC) that occured more often in 2015 than in 2016, and the last one was fixed by replacing the HVopto board during the maintenance period at the beginning of 2016. Seven channels started to be unstable only in 2016.

\begin{figure}[t]
\centering
\includegraphics[width=0.495\textwidth]{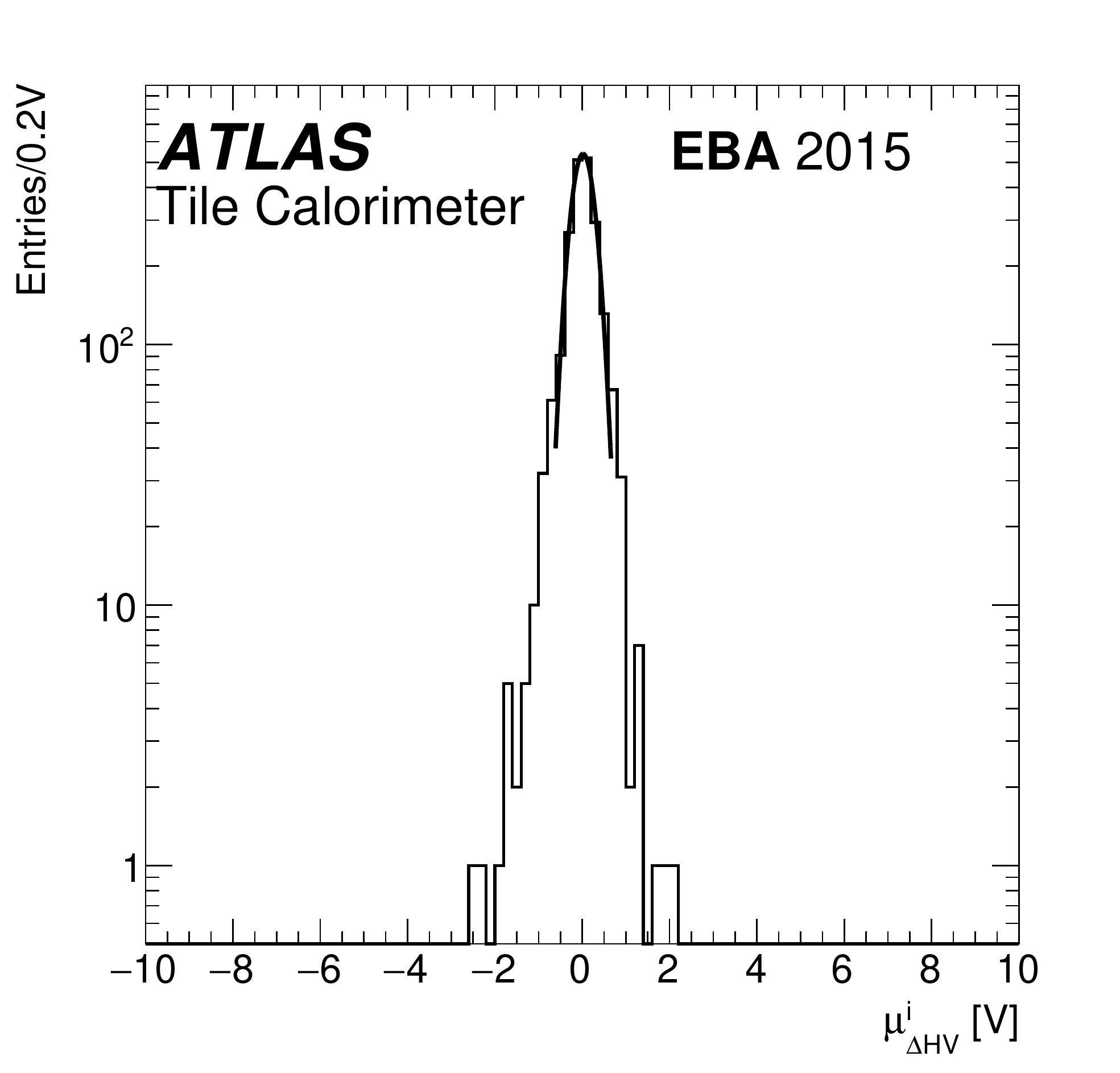}
%\caption{Distribution of $\mu_i$, defined in Subsection~\ref{subsec-5.2},  for all the channels of the EBA partition. The fitted Gaussian function is drawn and the corresponding parameters values reported.}
%\label{fig-9}       % Give a unique label
\caption{Distribution of \mudhvi, defined in Section~\ref{subsec:unstable},  for all the stable channels of the EBA partition obtained analysing 2015 data. 
%The fitted Gaussian function is drawn and the corresponding parameters values reported.
}
\label{fig:mui}       % Give a unique label
\end{figure}

Out of the 9569 (9595) stable channels identified in 2015 (2016), 88 (66) channels show an offset with respect to nominal value. They fail Eq.~(\ref{eq:dif}) condition and verify
\begin{equation}
%\mid  \mu_{part} -  \mu_i \mid \geq 5 \sigma_{part} 
\mid  \mudhvp-\mudhvi \mid \geq 5 \times \sigdhvp
\label{eq:off}                                         
\end{equation}
In Eq.~(\ref{eq:off}) \mudhvp{} and \sigdhvp{} are, respectively, the mean values and the sigmas of Gaussian function fits of the \mudhvi{} distributions obtained for the modules of the different partitions in the two considered run periods. Figure~\ref{fig:mui} shows as example the distribution of the \mudhvi{} obtained for the stable channels of the EBA partition  in 2015. In 2015 only three channels have a large offset with
\begin{equation}
%\mid  \mu_{part} -  \mu_i \mid \geq 5 \sigma_{part} 
\mid  \mudhvp -  \mudhvi \mid \geq 5\;\mathrm{V}
\label{eq:largeoff}                                         
\end{equation} 
%\textcolor{red}{In 2015, one channel shows a very broad \DeltaHV{} distribution: the values range in an interval of hundreds Volts.}
In 2016, among these three channels, one remained stable with the same large offset of $\approx10$~V while the two others became unstable. It must be noted that whatever the offset is, as long as the value is stable it does not affect the performances of the energy measurement. Indeed, the high voltage values are not used in the energy computation, the energy scale being set by the Cesium calibration~\cite{Cesium}. However, a large offset may be the hint of a serious problem in the regulation loop and it is useful for the maintenance.

%The total number of problematic channels, the ones showing a very broad distribution or instability or a large offset,
% (\ref{eq:dif}) and (\ref{eq:largeoff}) 
The total number of problematic channels, unstable or with a large offset,
was 27 in 2015 and 19 in 2016. They correspond to 0.3\% and 0.2\% of the investigated channels in the two periods. 

%\begin{table}[h]
%\centering
%\caption{Number of channels in each category for each year}   
%\label{tab:results}       % Give a unique label
% For LaTeX tables you can use 
%\begin{tabular}{	|c|c|c|} 
%\hline
%Year& 2015&2016   \\ \hline  \hline
%Analysed channels  & 9786   & 9748\\ \hline
%Good channels &  9569 &  9595 \\ \hline 
%Small offset & 85 &  65  \\ \hline  
%Large offset &  4 &  1\\ \hline  
%Bad shape &  78 &  45   \\ \hline 
%Really unstable &  24 &  18   \\ \hline  
%\end{tabular}
%\end{center}
%\end{table}

\subsection{Analysis of unstable channels using the Laser system}
\label{subsec:diagnosis}
%\begin{figure}[h]
%\centering
%\includegraphics[width=11cm,clip]{statusc.pdf}
%\caption{Status of the Tile Calorimeter HV channels in 2012. The HV channels that were working properly during the full year are in green. The yellow ones show large offsets. The red channels were unstable at least once during the year.  The blue ones had large offsets and were also unstable. The black rows represent channels in reduce HV-mode (see the text).}
%\label{fig:run1-status}       % Give a unique label
%\end{figure}
%The status of the Tile Calorimeter HV channels in 2012 is shown in Fig.~\ref{fig:run1-status}. In the scatter plots, the HV channels that were working properly during the full year are in green. The yellow ones show large offsets. The red channels were unstable at least once during the year.  The blue ones had large offsets and were also unstable. The channels in reduced HV-mode (see Subsection~\ref{subsec:HVopto}), for which no HV data was recoverable for at least 2 days, appear in black in the figure . 

Two different sources of instability can affect the channels: an individual failure of the regulation loop (usually from the opto-coupler or the 100 M$\Omega$ HV resistor) or of the monitoring part  (HV resistor or multiplexer or ADC). In the case of the regulation loop failure, {\HVmeas} is unstable because {\HVout} is really unstable, and this affects the stability of the measurement of the energy deposited in the calorimeter cells. In the case of the monitoring part failure, only {\HVmeas} is unstable while {\HVout} is stable and there is no effect on the energy measurement. The two sources of instability can be disentangled comparing the behaviours as a function of time of the gains of the PMTs determined using the values of the high voltage applied to the PMT ({\GHV}, computed from \HVmeas) and the one determined using the Laser calibration system~\cite{laser_note} ({\GLaser}, sensitive to \HVout):

\begin{equation}
\frac{\Delta\GHV(t)} {\GHV(\tref)} = \frac{\GHV(t)-\GHV(\tref)} {\GHV(\tref)} = \frac{ \left[\HVmeas(t)\right]^{\beta^i}} {\left[ \HVmeas(\tref)\right]^{\beta^i}} -  1 
\label{eq:gain_variation}                                         
\end{equation}

%where $G^i_{ref}$, the gain reference value corresponding to $HV^i_{ref}$ obtained  averaging the high voltage measurements over 10 minutes during the Laser run used to determine the $G^i_{LASER}$. 
%Figure~\ref{fig-12} shows the distribution of the measured $\beta$ values of the PMTs of TileCal. 
%The mean value of the measured $\beta$ distribution is 7.0 and the RMS is 0.2. Using Eq. (\ref{eq:gain_variation}) one obtains that a variation of the HV of 1 V corresponds roughly to a gain variation of about 1\%.
obtained using Eq. (\ref{eq:G}) and assuming $\HVmeas=\HVout$  and

\begin{equation}
\frac{\Delta\GLaser(t)} {\GLaser(\tref)} = \frac{\GLaser (t)-\GLaser(\tref)} {\GLaser(\tref)}~~ .
\label{eq:gain_variation_laser}                                         
\end{equation}

In the two cases, the reference values were determined in the first day of each of the two considered periods.

If $\HVmeas=\HVout$ and there is no specific problem with the Laser calibration, the two gain variations (on \GHV{} and \GLaser{}) should be equal. If a difference is observed, it means either that $\HVmeas\neq\HVout$ or that the gain computed by the Laser calibration is affected by other effects.

%The \HVref{} values were obtained averaging the high voltage measurements performed over 10 minutes during the Laser run used to determine the reference value of $G^i_{Laser}$. The uncertainty of the $G^i$ ($G^i_{Laser}$) measurements is 0.09\%~\cite{Valery:1743602}  (0.3\%~\cite{laser_note}). 
%The mean value of the measured $\beta$ distribution is 7.0 and the RMS is 0.2. Using Eq. (\ref{eq:gain_variation}) one obtains that a variation of the high voltage of 1 V corresponds roughly to a gain variation of ~\%.
%The uncertainty on the determinations of $G^i$ was estimated using a stable channels (see for example Fig.~\ref{fig:evolution}). The resolution on the $G^i$ measurements, 0.09\%, corresponds to the RMS of the distribution obtained projecting the scatter-plot points along the y-axis. The uncertainties of the $G^i_{Laser}$ measurements is 0.3\%~\cite{laser_note}. 

%The uncertainty on the determinations of $G^i$ was estimated studying the gain variation as a function of time in the case of stable channels. The obtained behaviour for the PMT EBC16\_13 is shown on the Fig.~\ref{fig-13}. The resolution on the $G^i$ measurements, corresponding to the RMS of the y-axis projected distribution, is 0.09\%. The corresponding error on  $G^i_{LASER}$ is 0.3~\%~\cite{laser_note}. 

\begin{figure}[t]
\centering
\includegraphics[width=0.495\textwidth]{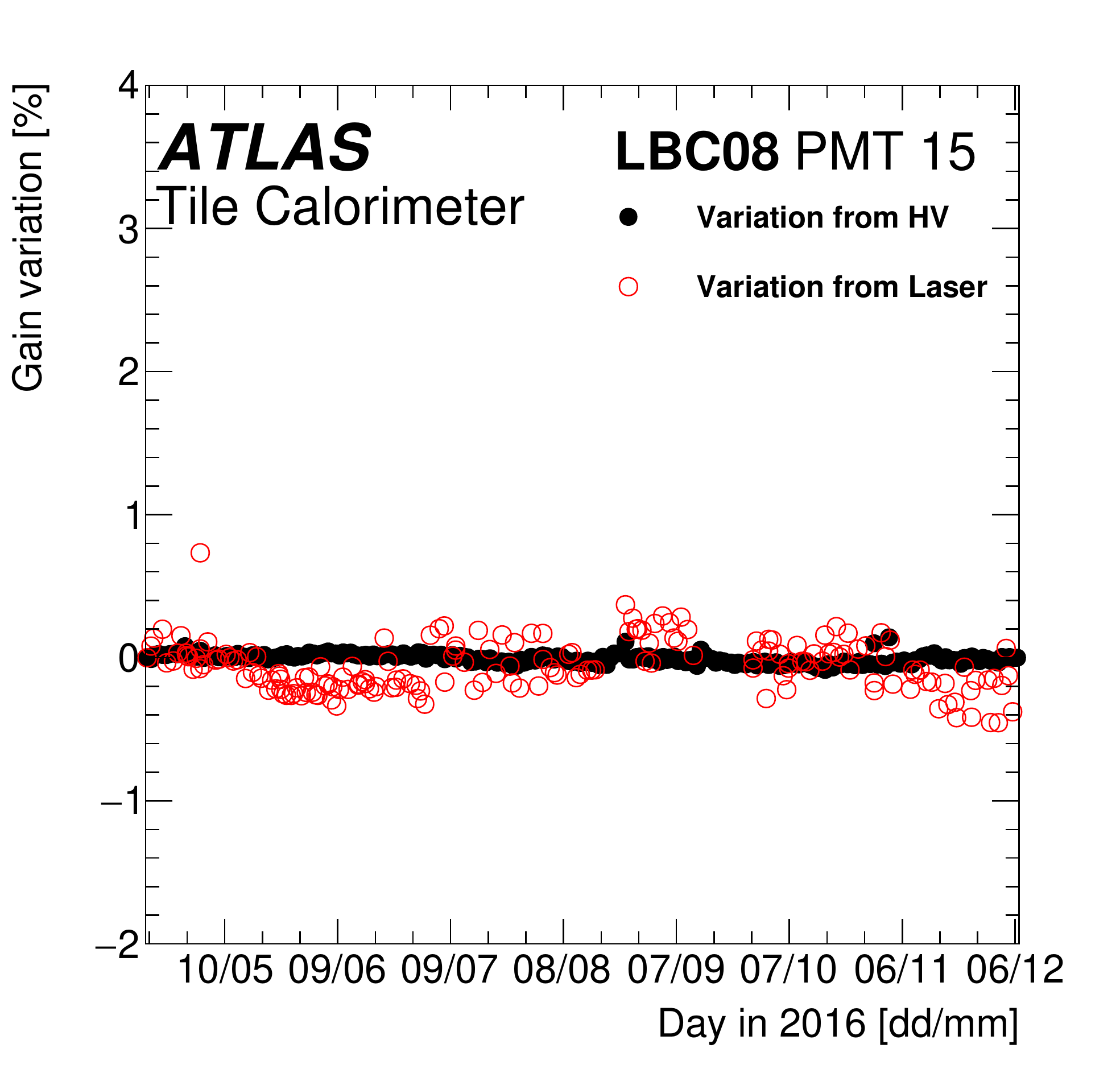}
\includegraphics[width=0.495\textwidth]{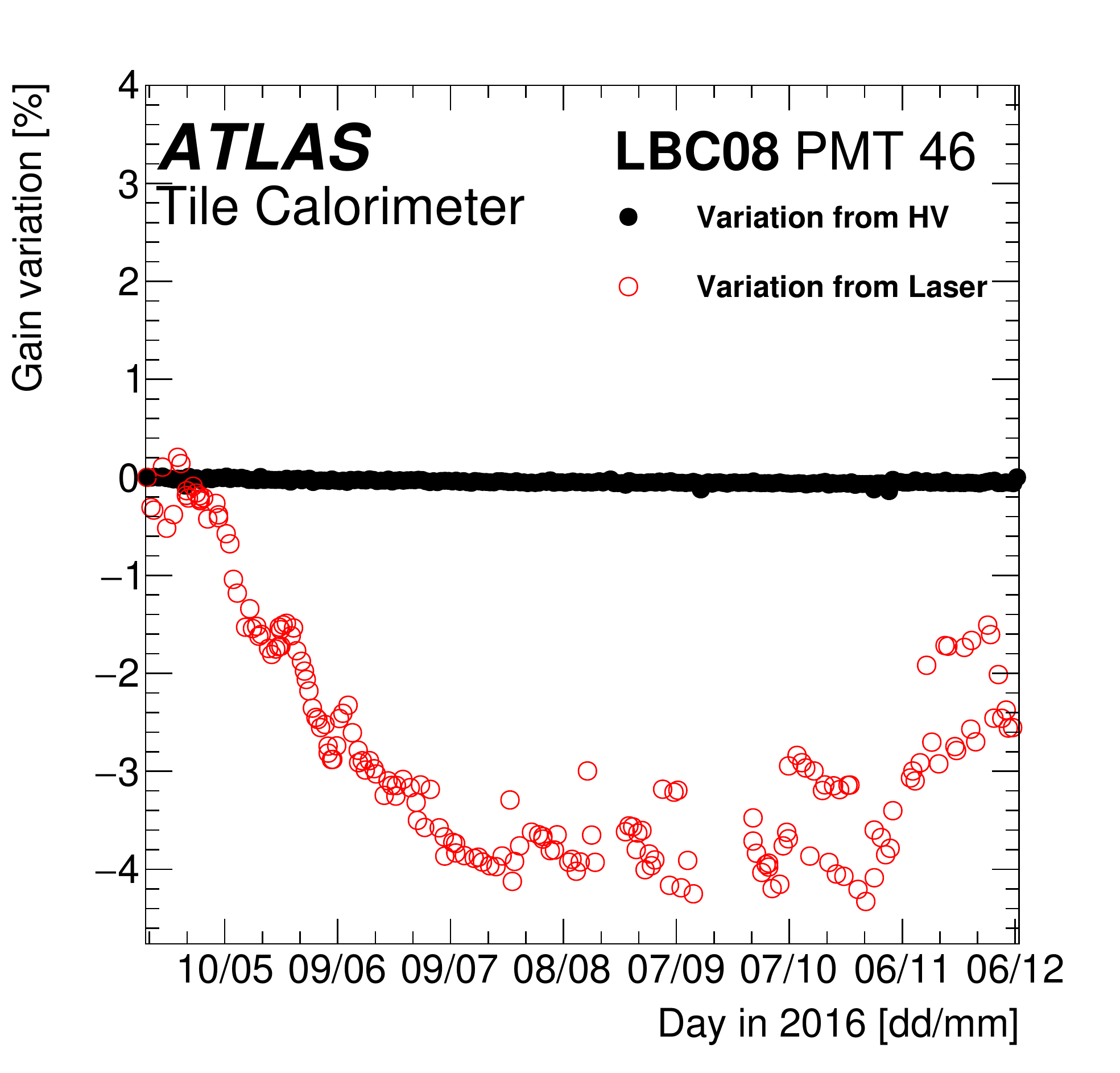}
\caption{Comparison of the PMT gain variation measured by the Laser and computed using the monitored high voltage {\HVmeas} in the case of the stable channels PMT 15 of the module LBC08 (left) and  PMT 46 of the module LBC08 (right). By definition, the first point is set to zero.}
\label{fig:stable}       % Give a unique label
\end{figure}

\begin{figure}[t]
\begin{center}
\includegraphics[width=0.495\textwidth]{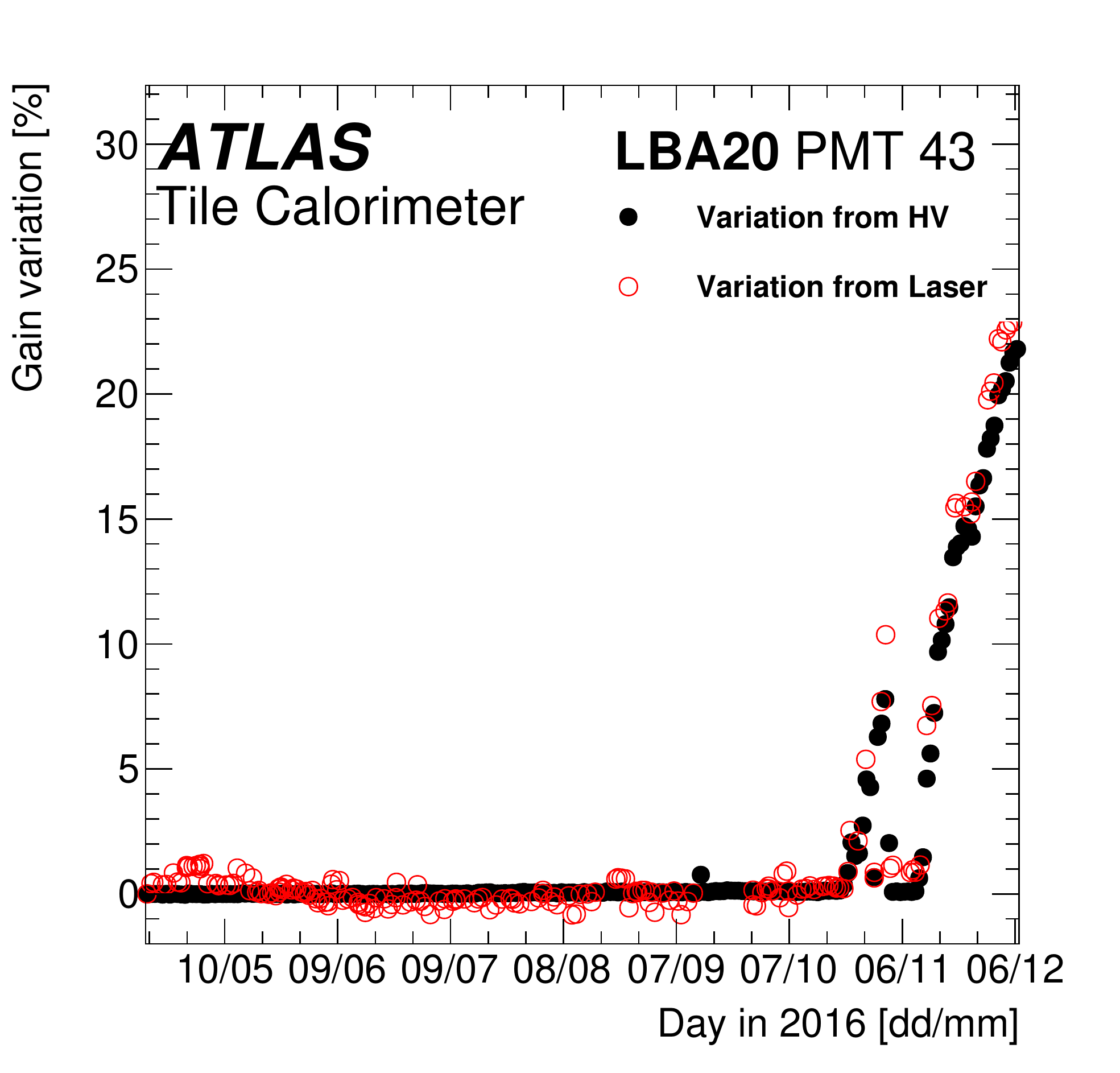}
\includegraphics[width=0.495\textwidth]{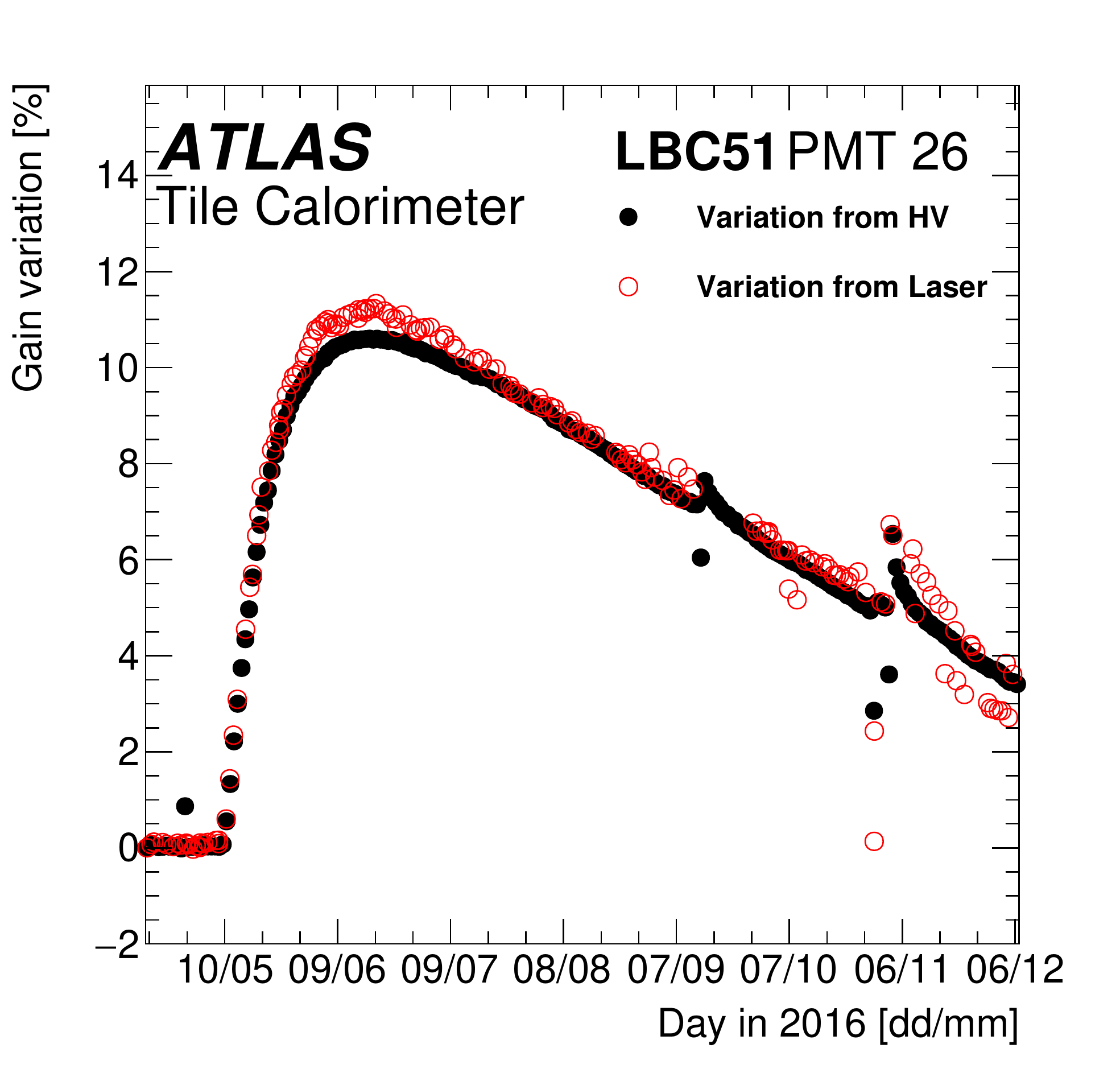}
\caption{Comparison of the PMT gain variation measured by the Laser and computed using the monitored high voltage {\HVmeas} in the case of channels with faults in the regulation loop systems: PMT 43 of the module LBA20 (left) and  PMT 26 of the module LBC51 (right). By definition, the first points are set to zero.}
\label{fig:unstablestable}
\end{center}
\end{figure}

In the following the results obtained analyzing 2016 data are discussed. Unfortunately, the gain of the PMTs obtained using Eq.~(\ref{eq:gain_variation_laser}) is also affected by the PMT illumination. In the considered period, due to the high luminosity of the LHC collider, a decrease of the gain between 2 and 5\% was observed for most of the channels~\cite{laser_note}. This makes difficult to correlate gain variations on {\GHV} and {\GLaser} for instabilities smaller than approximately 5~V. An example of comparison can be seen for two channels of LBC08
in Fig.~\ref{fig:stable}. The PMT 15 on the left is very stable and does not suffer from any luminosity effect: since it corresponds to a cell far from the collision point, the gain variation measured by the Laser is compatible with the one obtained from the {\HVmeas}. PMT 46 on the right of the figure, corresponding to a cell closer to the collision point, shows stable values of \GHV{} but \GLaser{} is affected by the variation of the gain due to luminosity. The effect produces a discrepancy of the two gains behaviours as a function of time, although {\HVout} is stable and well measured.
Among the 18 unstable channels, 13 of them exhibit a rather clear correlation between the observed variation of {\GHV}  and the PMT gain variation measured by the Laser, thus pointing at problems linked to the regulation loop and therefore real instabilities of {\HVout}. As an example Fig.~\ref{fig:unstablestable} shows the behaviour of two really unstable channels: the instabilities measured with \HVmeas{} are compatible with the gain variation measured by the Laser, the luminosity effect being negligible (cells with low illumination).

\begin{figure}[t]
\centering
\includegraphics[width=0.495\textwidth]{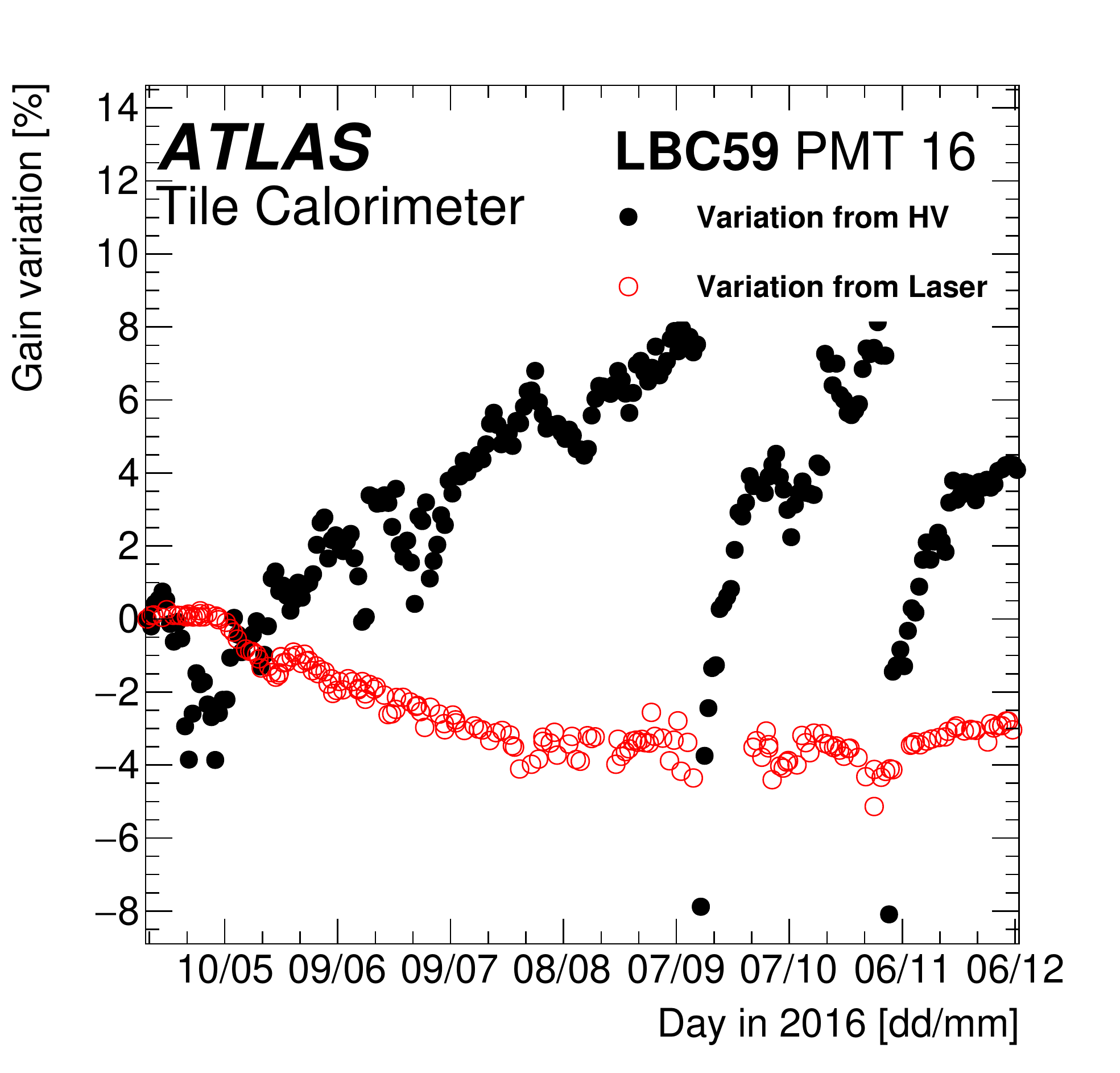}
\caption{Comparison of the PMT gain variation measured by the Laser and computed using the monitored high voltage {\HVmeas} in the case of  a channel with faults in the monitoring system. By definition, the first point is set to zero.}
\label{fig:unstableunstable}       % Give a unique label
\end{figure}

The other five channels show different behaviours of the gains {\GHV} and {\GLaser}. One example is shown in Fig.~\ref{fig:unstableunstable}, where one can clearly see the luminosity effect on {\GLaser} (decrease of about 4\%) and a completely different behaviour for {\GHV}.
%The results show  that for these channels the reading of the high voltage performed trough a separate electronic loop is not working properly.
For these channels, the monitoring of the applied high voltage is not working properly but the value of \HVout{} seems to be stable since the gain variation measured by the Laser is compatible with the normal decrease due to luminosity.

From this analysis one can conclude that the number of really problematic HV channels in the TileCal detector is less than 0.2\%. Moreover, since the calorimeter cells are read-out by two PMTs, the impact of the studied effects on the jet reconstruction is negligible. In particular, no cell is suffering from instabilities of the two PMTs. Similar results were found analysing 2012 data~\cite{Valery:1743602}. 

\clearpage
%-------------------------------------------------------------------------------
\section{Conclusions}
\label{conclusions}
%-------------------------------------------------------------------------------
%\textcolor{red}{To be improved}

The high voltage  system of the TileCal calorimeter of the ATLAS experiment allows to set and monitor the  high voltage value of each of the 9852 PMTs of the sub-detector. In this article the hardware and the online software of the system have been described. 
%According to the design, the sensitivity to temparature was found to be smaller than 0.2~V/$^\circ$C. Keeping the humidity below 60\% no HV trip was detected. 
The analysis of the data collected in 2015 and 2016 show very few problematic channels. Only 0.2~\% of the PMTs show a gain instability larger than 0.5\%.  
\printbibliography
% If you want to use the traditional BibTeX you need to use the syntax below.
%\bibliographystyle{bibtex/bst/atlasBibStyleWoTitle}
%\bibliography{atlas-document,bibtex/bib/ATLAS}

%-------------------------------------------------------------------------------
%\clearpage
%\appendix
%\part*{Auxiliary material}
%\addcontentsline{toc}{part}{Auxiliary material}
%-------------------------------------------------------------------------------

%In an ATLAS paper, auxiliary plots and tables that are supposed to be made public 
%should be collected in an appendix that has the title \enquote{Auxiliary material}.
%This appendix should be printed after the Bibliography.
%At the end of the paper approval procedure,
%this information should be split into a separate document -- see \texttt{atlas-auxmat.tex}.

\end{document}